\documentclass[printer]{aa}
\usepackage{rotating}
\usepackage{upgreek}
\usepackage{txfonts}
\usepackage{booktabs}
\usepackage{graphicx}
\usepackage{graphics}
\usepackage{natbib}
\usepackage{float}
\usepackage{placeins}
\usepackage{multirow}
\usepackage{lscape}
\usepackage{pdflscape}
\usepackage{gensymb}
\usepackage[export]{adjustbox}
\usepackage{longtable}
\usepackage{supertabular,booktabs}
\usepackage{auto-pst-pdf}
\bibpunct{(}{)}{;}{a}{}{,} 
\def\new#1{\textbf{#1}}
\begin{document}
\newcommand{\msun}{\mbox{M$_{\odot}$}}
\newcommand{\rsun}{\mbox{R$_{\odot}$}}
\newcommand{\lsun}{\mbox{L$_{\odot}$}}
\newcommand{\tsun}{\mbox{T$_{\odot}$}}
\title{The evolutionary status of dense cores in the NGC~1333 IRAS~4 star-forming region}

\author{E.~Koumpia\inst{\ref{inst1},\ref{inst2},\ref{inst9}}, F.F.S.~van der Tak\inst{\ref{inst1},\ref{inst2}}, W.~Kwon\inst{\ref{inst3},\ref{inst8}}, J.J.~Tobin\inst{\ref{inst4},\ref{inst5}}, G.A.~Fuller\inst{\ref{inst6}}, R.~Plume\inst{\ref{inst7}}}

\institute{SRON Netherlands Institute for Space Research, Landleven 12, 9747 AD Groningen, The Netherlands\label{inst1}
\and
Kapteyn Institute, University of Groningen, Landleven 12, 9747 AD Groningen, The Netherlands\label{inst2}
\and 
Korea Astronomy and Space Science Institute, 776, Daedeok-Daero, Yuseong-gu, Daejeon, 34055, Korea\label{inst3}
\and
National Radio Astronomy Observatory, Charlottesville, VA 22903, USA\label{inst4}
\and 
Leiden Observatory, Leiden University, P.O. Box 9513, 2300-RA Leiden, The Netherlands\label{inst5} 
\and
Jodrell Bank Centre for Astrophysics \& UK ALMA Regional Centre Node, School of Physics \&
Astronomy, The University of Manchester, Manchester, M13 9PL, UK\label{inst6}
\and
Department of Physics \& Astronomy and Institute for Space Imaging Sciences, University of Calgary, Calgary, AB T2N 1N4, Canada\label{inst7}
\and
Korea University of Science and Technology, 217 Gajeong-ro, Yuseong-gu, Daejeon 34113, Korea\label{inst8}
\and \email{ev.koumpia@gmail.com}\label{inst9}
}
\date{Received date/Accepted date}

\begin{abstract}
{Protostellar evolution, following the formation of the protostar is becoming reasonably well characterized, but the evolution from a prestellar core to a protostar is not well known, although the first hydrostatic core (FHSC) must be a pivotal step.} 
{NGC~1333 -- IRAS~4C is a potentially very young object,
that we can directly compare with the nearby Class 0 objects IRAS~4A and IRAS~4B. Observational constraints are provided by spectral imaging from the JCMT Spectral Legacy Survey (330--373 GHz). We present integrated intensity and velocity maps of several species, including CO, H$_{2}$CO and CH$_{3}$OH. CARMA observations provide us with additional information to distinguish IRAS~4C from other evolutionary stages.} 
{The velocity of an observed outflow, the degree of CO depletion, the deuterium fractionation of [DCO$^{+}$]/[HCO$^{+}$] and gas kinetic temperatures are observational signatures that we present.} 
{We report differences between the three sources in four aspects: a) the kinetic temperature as probed using the H$_{2}$CO lines is much lower towards IRAS~4C than the other two sources, b) the line profiles of the detected species show strong outflow activity towards IRAS~4A and IRAS~4B but not towards IRAS~4C, c) the HCN/HNC is $<$ 1 towards IRAS~4C, which confirms the cold nature of the source, d) the degree of CO depletion and the deuteration are the lowest towards the warmest of the sources, IRAS~4B.}
{IRAS~4C seems to be in a different evolutionary state than the IRAS~4A and IRAS~4B sources. We can probably exclude the FHSC stage due to the relatively low L$_{smm}$/L$_{bol}$ ($\sim$6~\%) and we investigate the earliest accretion phase of Class~0 stage and the transition between Class~0 to Class~I. Our results do not show a consistent scenario for either case, with the major issue being the absence of outflow activity and the cold nature of IRAS~4C. The number of FHSC candidates in Perseus is $\sim$10 times higher than current models predict, which suggests that the lifespan of these objects is $\ge$ 10$^{3}$ years, possibly due to an accretion rate lower than 4$\times$10$^{-5}$$~\msun$/yr.}

\end{abstract}


\titlerunning{The evolutionary status of dense cores in the NGC~1333 IRAS~4 region} 
\authorrunning{Koumpia et al.} 
\maketitle


\section{Introduction} 

Stars form by gravitational collapse of dense cores in molecular clouds. In order to understand the origin 
of stellar masses, multiple systems and outflows, it is necessary to understand the formation and evolution 
of dense cores. Representing the earliest phase of star formation, both prestellar and protostellar 
cores have been observed and studied using large (sub--) millimeter telescopes \citep[e.g. JCMT;][]{Enoch06,Kempen06}, 
infrared \citep[e.g. Spitzer;][]{Evans03,Evans09,Jorgensen06,Hatchell2005,Young04} and interferometers 
\citep[e.g. CARMA;][]{Enoch10,DiFrancesco01}. While the basic evolutionary ordering of protostellar objects seems firmly established \citep[e.g.;][]{Andre2000}, we still do not have a clear view of the evolutionary process that turns a prestellar core into a protostar \citep{Bergin07,Evans09}. 

The transition from pre--stellar to protostellar cores is predicted to be the first hydrostatic core, FHSC \citep{Larson69}, which represents the phase after the collapse of the parent core and before the formation of a protostar. This object is characterized by a very short lifetime \citep[$\sim$1000 years;][]{Machida2008}. Recent studies have provided a few candidates including Per--bolo 58, L1448--IRS2E, LDN 1451--mm, B1--bN, B1--bS and Per--bolo 45\citep{Enoch10, Chen2010, Pineda2011, Pezzuto2012, Schnee2012}. Statistically, using 

\begin{equation}
N_{FHSC} = N_{CLASS0}\frac{\uptau_{FHSC}}{\uptau_{CLASS0}}
\label{FHSC}
\end{equation}

and assuming a lifespan of $\sim$10$^{3}$ yrs for FHSC and 2.2$\sim$10$^{5}$ yrs for Class~0 objects \citep{Enoch09}, while there are about 27 Class~0 objects in Perseus \citep{Enoch09}, the predicted number of FHSC in Perseus is $\le$ 0.2. When adopting a lifetime of $\sim$10$^{4}$ yr for Class~0 objects based on outflow kinematic timescale \citep{Machida2013}, the predicted number of FHSC in Perseus is about 5, which is closer to the observed number.

Another very early stage of low--mass star formation is the very low luminosity object (VeLLO) which is characterized by an internal luminosity of $\le$0.1$\lsun$. \citet{Dunham2008} report 15 VeLLO candidates, while even fewer have been studied in detail \citep[e.g. L1014--IRS, L1521F--IR, IRAM 04191+1522;][]{Young04,Bourke2006,Dunham2006}. VeLLOs have been associated with a brown dwarf progenitor \citep{Lee2013} or a very low mass Class~0 protostar with low accretion, but their exact nature is still uncertain. 

The NGC~1333 region in the Perseus molecular cloud is an excellent laboratory to study early stages of low--mass star formation. NGC~1333 is a nearby \citep[D=235~pc;][]{Hirota2008} and young \citep[$<$~1~My;][]{Gutermuth2008} star forming 
region. It is a part of the Perseus OB2 molecular cloud complex which contains a large number of young stellar objects (YSOs). It hosts about 50 YSOs and 36 Herbig--Haro objects. 

This work focuses on three YSOs in NGC 1333: IRAS~4A, IRAS~4B and IRAS~4C. While IRAS~4A and IRAS~4B are well studied Class 0 objects, the nature of IRAS~4C is still under debate. Previous studies, including \citet{Enoch09} and \citet{Sadavoy2014} classified IRAS~4C also as a Class 0 YSO. Both IRAS~4A and IRAS~4B are found to be binaries. \citet{Sandell2001} and \citet{DiFrancesco01} resolved the IRAS~4B/IRAS~4B~II binary using JCMT and PdBI reporting the fainter binary companion IRAS~4B~II at $\sim$10 $\arcsec$ to the east of IRAS~4B. IRAS~4B~II has been called IRAS~4C as well \citep[e.g.;][]{Looney2000} but IRAS~4C is generally used as the source in our work which is $\sim$40$\arcsec$ east--northeast of IRAS~4A \citep[e.g.;][]{Smith2000}. 

Interferometric observations of ammonia towards all three sources were briefly reported in \citet{Wootten1995}, where only the emission from IRAS~4A was clearly associated with outflowing warm gas. In the same study IRAS~4C does not show that clear association with the structures located near the protostellar environment. Interferometric observations of ammonia and of NH$_{2}$D towards IRAS~4A and IRAS~4C were reported in \citet{Shah2001} and indicated colder conditions towards IRAS~4C.  

This paper uses JCMT (James Clerk Maxwell Telescope) molecular line emission and CARMA (Combined Array for Research in Millimeter-wave Astronomy) dust continuum observations to clarify the evolutionary status of IRAS~4C, 
using the nearby IRAS~4A and IRAS~4B sources as comparison standards and study the physical and chemical structure that characterizes these very early evolutionary stages. Our study of IRAS~4C is unique since it provides information about the spatial and velocity structure of many molecules while it has previously been studied mainly photometrically \citep[e.g.;][]{Dunham2008,Enoch09,Sadavoy2014}.

In order to understand the nature of IRAS~4C, it is crucial to obtain the properties 
of the dust and gas 
that are present in its envelope and
outflow. One of the observational signatures that distinguish a FHSC from 
a typical protostar is the velocity of the molecular outflow. 
According to models, a FHSC should be able to drive an early outflow which is expected to be weak and 
of low velocity V $\sim 3$ km~s$^{-1}$ \citep{Machida2008,Tomida2010}, 
while the typical speed of a protostellar outflow is 10--20 km~s$^{-1}$ \citep{Arce06}. 
Thus, measuring the velocity of
an observed outflow could test the evolutionary stage of IRAS~4C. Water masers could indicate an outflow activity but several interferometric surveys have not detected associated maser emission near IRAS~4C \citep{Rodriguez2002,Furuya2003,Park2007}.

Another observational tool is the deuterium fractionation (e.g. [N$_{2}$D$^{+}$]/[N$_{2}$H$^{+}$]), 
which can be used as a chemical clock \citep{Belloche06,Fontani2011,Fontani14}. Deuterated species are enhanced in environments of low temperatures (T$<$ 20~K) where CO is depleted \citep{Millar1989}. The coldest (i.e. the youngest) objects are characterised by the largest deuterium fractionation \citep{Crapsi2005a}.

Although there is not a single observational signature that determines the evolutionary stage of such young objects, taken together, the observational properties listed above provide strong tools to distinguish the evolutionary status of IRAS~4C compared to IRAS~4A or IRAS~4B. 

\section{Observations and data reduction} 

\subsection{JCMT}

Spectral maps of the NGC~1333~IRAS~4 region were taken as a part of the 
JCMT Spectral Legacy Survey \citep[SLS;][]{Plume2007} with the 16--element
Heterodyne Array Receiver Programme B (HARP--B) and the
Auto--Correlation Spectral Imaging System (ACSIS) at
the James Clerk Maxwell Telescope (JCMT\footnote{The James Clerk Maxwell Telescope has historically been operated by the Joint Astronomy Centre on behalf of the Science and Technology Facilities Council of the United Kingdom, the National Research Council of Canada and the Netherlands Organisation for Scientific Research.}) on Mauna Kea, Hawaii. 
HARP--B consists of 16 pixels providing high-resolution (1~MHz, $\sim$ 1~km~s$^{-1}$) maps of a 
2\arcmin$\times$2\arcmin~field. The original frequency coverage of 330--360 GHz of the survey was complemented by HARP maps 
at the window of the higher frequencies (360--373 GHz) as a result of additional proposals between 2007 and 2010. 
The observations were performed in a jiggle position switch mode to create maps of a 2\arcmin$\times$2\arcmin~area with 
pixels spaced by 7.5$\arcsec$. Our maps were centered at RA = 03:29:11.3, Dec = 31:13:19.5 (J2000). The spectra were taken 
using an off--position at RA = 03:30:21.0, Dec = 31:13:19.5 (J2000) which is about 15\arcmin~east of the field center. The angular resolution of the JCMT is 
$\sim 15$$\arcsec$ at 345~GHz which is equivalent to $\sim 3000$ AU at the distance of NGC~1333~IRAS~4 \citep{Choi2004}. The beam efficiency is 0.63 \citep{Buckle2009}. 

The raw time series files were reduced semi--manually using standard 
procedures from the Starlink software package \footnote{See http://starlink.eao.hawaii.edu/starlink} in combination with 
the ORAC Data Reduction pipeline (ORAC-DR). Each scan file was inspected and 
corrected for bad receptors, baseline subtraction and spike removal, 
before converting time series frequencies to 3--D data cubes. 
The pipeline output was further checked and corrected for remaining baseline 
issues, high rms noise level cubes or position issues using specific tasks 
from Starlink. The overall noisy band edges were ignored, allowing 
$\sim$~0.85 GHz of bandwidth per frequency block to be used. The overall rms noise level is ranging between 0.005 and 0.05~K for the majority of the cubes 
with detected lines. 

\subsection{CARMA}

In order to study the small--scale structure of IRAS~4C (RA = 03:29:13.64, Dec = 31:13:57.5 (J2000)), including dust properties, we obtained single pointing interferometric CARMA \footnote{Combined Array for Research in Millimeter-wave Astronomy; See https://www.mmarray.org} observations at 1.3~mm, between April and July 2014. The 15--element array excluding eight of 3.5-meter antennas consists of nine 6.1--meter diameter telescopes and six 10.4--meter diameter telescopes with a primary beam at 230 GHz of 47\arcsec and~28\arcsec~respectively.

\begin{figure*}[ht]
\includegraphics[width=\hsize]{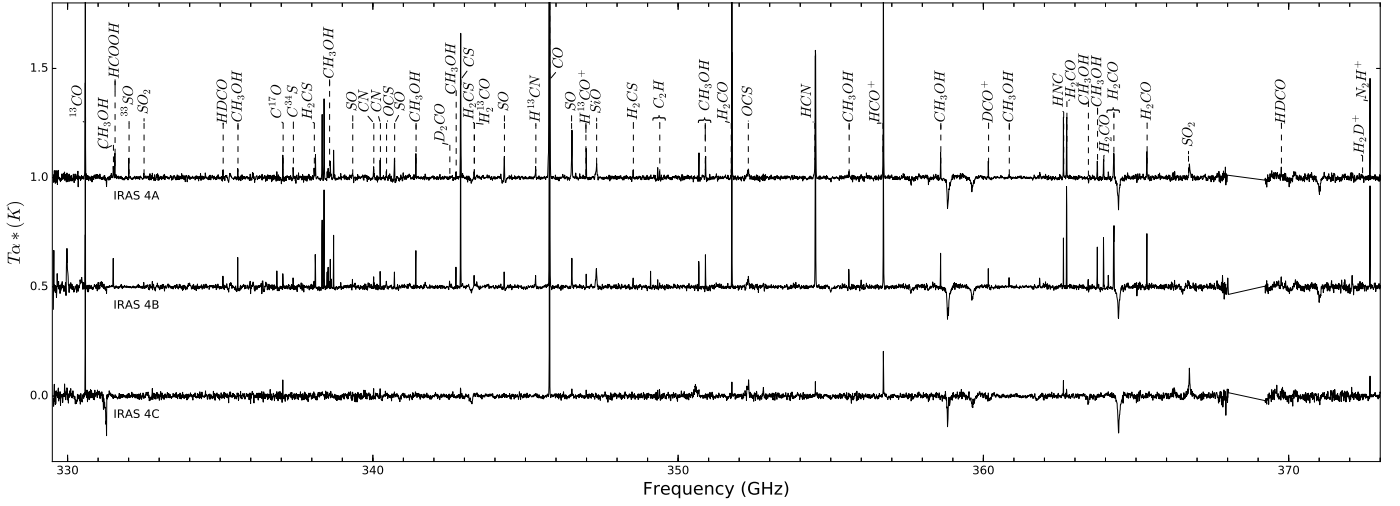}
\caption{Main line detections of the spectra extracted for the 3 positions of NGC~1333: IRAS~4A, IRAS~4B and IRAS~4C covering the full JCMT range 330--373 GHz. The gap seen at 368--369.5~GHz is due to poor atmospheric transmission. The spectra are plotted with a vertical offset at 0.3, 0.7 and 1.1 intensity levels (K) for easier comparison. The coordinates of the three sources IRAS~4A (RA = 03:29:10.51, Dec = 31:13:31.4 (J2000)), IRAS~4B (RA = 03:29:12.01, Dec = 31:13:08.0 (J2000)) and IRAS~4C (RA = 03:29:13.64, Dec = 31:13:57.5 (J2000)) are taken from \citet{Sandell2001}.}
\label{fig:speccomp}
\end{figure*}


In addition to the dust continuum emission, we observed the following molecular transitions: CO 2-1 (230.5 GHz), $^{13}$CO 2-1 (220.4 GHz), C$^{18}$O 2-1 (219.6 GHz) and N$_{2}$D$^{+}$ (231.3 GHz), which are tracers of an envelope and/or outflow. These observations were performed in a dual polarization mode that provided us with a better sensitivity in D configuration. For the CO isotope lines, a 31 MHz bandwidth was used with 3--bit sample mode, which provides a good velocity resolution ($\sim0.13$ km~s$^{-1}$) and coverage ($\sim40$ km~s$^{-1}$), in order to trace the kinematics of the envelope as well as the outflow. These 4 molecular lines were observed simultaneously using 3 bands and the additional fourth band was set to the wide 500 MHz bandwidth for the continuum emission and calibration. The CARMA data are characterized by 2$\arcsec$ angular resolution ($\sim$500~AU at the target distance) and $\sim$15~mJy/beam sensitivity. Because the typical dense cores and envelope sizes are around 5000~AU \citep{Crapsi05}, our data resolve the structures well.

We reduced and edited the visibility data using the MIRIAD software package \citep{Sault95}. 

\section{Observational results} 

\subsection{Single dish line detections and morphology}
\label{sec:maps}

The JCMT data provide us with information about the spatial and velocity distributions of many different molecules. The majority of lines show clear emission towards the Class~0 IRAS~4A and IRAS~4B sources whereas fewer molecules show compact emission towards IRAS~4C (Figure~\ref{fig:speccomp}). 

\begin{figure}[h]
\includegraphics[scale=0.45, angle=270]{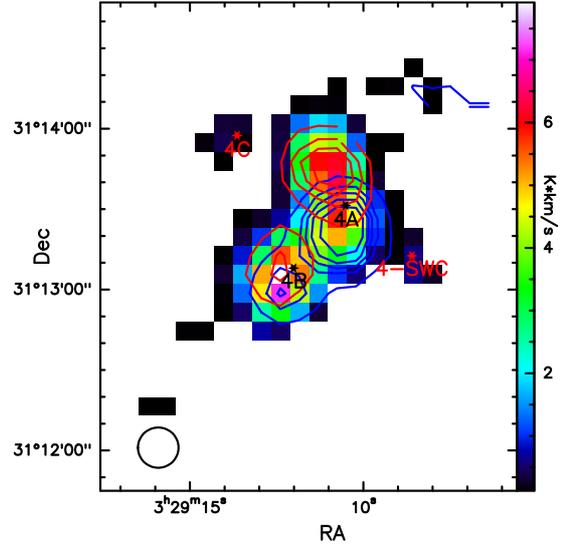}
\caption{Integrated intensity map (core; from $+$5 to $+$10~km~s$^{-1}$) of H$_{2}$CO 5$_{1,5}$--4$_{1,4}$ in colors, overplotted with its blue (from -5 to $+$5~km~s$^{-1}$) and red (from $+$10 to $+$17 ~km~s$^{-1}$) wing emission in blue and red contours respectively. H$_{2}$CO peaks towards IRAS~4B and shows a weak emission towards IRAS~4C but not connected to outflows. Note also the small peak to the southwest of IRAS~4A. The red and blue contour levels are set to 10, 30, 50, 70, 100 $\times$rms (rms: 0.02~K).}
\label{fig:h2co_spatial}
\end{figure}

Figure~\ref{fig:h2co_spatial} shows the H$_{2}$CO 5$_{1,5}$--4$_{1,4}$ emission as a representative example of the spatial distribution seen in all 3 sources. Emission from IRAS~4A and IRAS~4B and their surroundings is prominent with the strongest emission towards IRAS~4B. There is also significant emission towards the north of IRAS~4A\new{,} as a result of its strong outflow activity. The peak of the red wing emission (from $+$10 to $+$17~km~s$^{-1}$) of H$_{2}$CO is also at that position. Weak emission of H$_{2}$CO can be seen also towards IRAS~4C, which is more compact. The red and blue wings trace the red--shifted and blue--shifted outflow activity from both IRAS~4A and IRAS~4B but there is no such signature from IRAS~4C at those velocities. In the same plot a fourth source of emission is also revealed towards the southwest of IRAS~4A, which we call IRAS~4--SWC (southwest clump) in this work.  

The detected species (Table~\ref{JCMT_lines_1}--\ref{JCMT_lines_3}) are divided into two groups. One group is characterized by narrow line profiles (FWHM $\sim$ 1--5~km~s$^{-1}$) and compact spatial distribution ($\sim$15--20\arcsec) and is therefore suggested to arise in the quiescent envelope (e.g. C$^{17}$O, HCO$^{+}$, H$_{2}$CS and N$_{2}$H$^{+}$). The other group is characterized by broad line profiles (FWHM $\sim$ 8--16~km~s$^{-1}$) and extended emission ($\sim$40\arcsec -- 1\arcmin) and is therefore suggested to trace dynamical processes such as outflows (e.g. CO, HCN, CH$_{3}$OH and H$_{2}$CO). 

\begin{figure}[ht]
\begin{center}
\includegraphics[scale=0.4, angle=180]{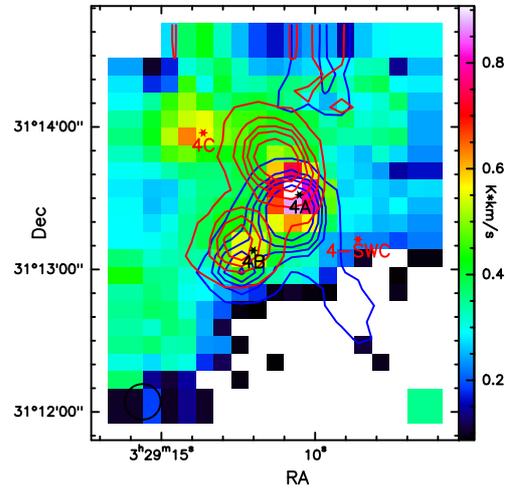}
\end{center}
\caption{Integrated intensity map (core; from $+$5 to $+$9~km~s$^{-1}$) of C$^{17}$O 3--2 in colors, overplotted with the HCN 4--3 contours tracing outflow activity, namely red wing (from $+$10 to $+$25 ~km~s$^{-1}$) in red contours and blue wing (from -10 to $+$4 ~km~s$^{-1}$) emission in blue contours. The plot shows the presence of the bipolar outflows driven by IRAS~4A and IRAS~4B Class~0 objects. We do not see such activity towards IRAS~4C. C$^{17}$O is stronger towards IRAS~4C than towards IRAS~4B. The red and blue contour levels are set to 10, 30, 50, 70 and 100 $\times$rms (rms: 0.02~K). The contours on the top are due to more noisy map towards the edges.}
\label{fig:c17o}
\end{figure}

Figure~\ref{fig:c17o} shows the C$^{17}$O integrated intensity map (from $+$5 to $+$9~km~s$^{-1}$) tracing the quiescent gas overplotted with the red (from $+$10 to $+$25~km~s$^{-1}$) and blue (from -10 to $+$4~km~s$^{-1}$) wings of the HCN 4-3 line profiles tracing the activity shifted from the LSR velocity (the signature of an outflow). 

\begin{figure}[ht]
\includegraphics[scale=0.4, angle=180]{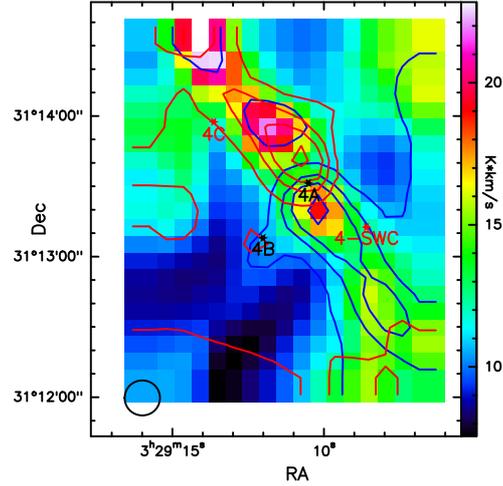}
\caption{Integrated intensity map (core; from $+$5 to $+$9~km~s$^{-1}$) of CO 3-2 in colors, overplotted with the red wing (from $+$10 to $+$25~km~s$^{-1}$) emission (red contours) and blue wing (from -10 to $+$4 ~km~s$^{-1}$) emission in blue contours. The plot clearly shows the strong bipolar outflow driven by IRAS~4A Class~0 object. The red and blue contours levels are set to 100, 300, 500 and 800 $\times$rms (rms: 0.02~K).}
\label{fig:CO_3_2}
\end{figure}

The IRAS~4A and IRAS~4B Class~0 objects both drive bipolar outflows first detected in CO and CS by \citet{Blake1995} and a few years later in millimeter transitions of SiO by \citet{Lefloch1998}. The strong bipolar collimated outflow driven by IRAS~4A can also be seen in Figure~\ref{fig:CO_3_2}, where IRAS~4B shows a compact outflow and IRAS~4C does not show a bipolar outflow, but a cone--like structure to the east of IRAS~4C. The continuum emission and the starting point of the cone show an east offset of 14\arcsec, and given the fact that IRAS~4C is the closest source it 
might be associated with it, but this is not clear from our observations. A water maser was reported by single--dish observations near IRAS~4C by \citet{Haschick1980}, a fact that could indicate the presence of an outflow. However, several following interferometric surveys did not detect associated maser emission at the reported position \citep{Rodriguez2002,Furuya2003,Park2007}. Figure 1 presented by \citet{Park2007}, shows that the previously reported H$_{2}$O maser is $\sim$25\arcsec to the north of IRAS~4A and west of IRAS~4C and follows the distribution of the well known outflow activity from IRAS~4A. This is an indication that this maser is not really connected to IRAS~4C.

Our CO maps are very similar to the ones presented by \citet{Yildiz2012}, where the 6--5 transition is also included. IRAS~4A is a binary system \citep{Looney2000} with a separation of 1.8$\arcsec$ (420~AU at a distance of 235~pc) and it is characterized by two outflows with different directions. The strongest outflow seen in CO is in the NE--SW direction, while other tracers such as H$_{2}$CO (Figure~\ref{fig:h2co_spatial}) show a more compact emission on the N--S axis. Similar findings are presented by \citet{Santangelo2015}. The binary nature of IRAS~4A could explain these two different outflow morphologies observed in different tracers.

\begin{figure}[ht]
\begin{center} 
\includegraphics[scale=0.4, angle=180]{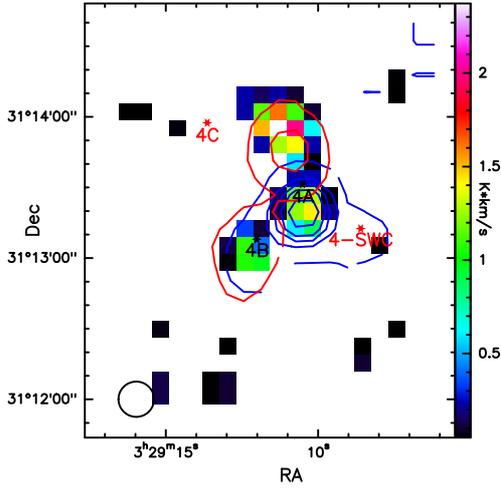}
\end{center}
\caption{Integrated intensity map (from -11 to $+$23 ~km~s$^{-1}$) of SiO 8--7 in colors, overplotted with the blue (from -5 to $+$5 ~km~s$^{-1}$) and red (from $+$9 to $+$15~km~s$^{-1}$) wing emission from CH$_{3}$OH 7$_{-1,7}$--6$_{-1,6}$ in blue and red contours respectively. The SiO peaks north of IRAS~4A close to the peak redshifted emission of CH$_{3}$OH and traces the shock activity in the region. No emission connected to IRAS~4C is observed. The red and blue contour levels are set to 5 30 50 70 100 and 150$\times$rms (rms: 0.02~K).}
\label{fig:sio}
\end{figure}

 Figure~\ref{fig:sio} shows the SiO integrated intensity map in the full range from $-$11 to $+$23~km~s$^{-1}$. The peak of this emission is northeast of IRAS~4A, following the red--shifted outflow activity from IRAS~4A as traced with methanol (CH$_{3}$OH, Figure~\ref{fig:sio}). The spatial distribution of SiO follows the same pattern as the spatial distribution of the extracted outflow activity from CH$_{3}$OH. SiO traces even more energetic processes such as shocks \citep{Duarte2014}. In these tracers IRAS~4C is absent, which may be an indication of less energetic outflow activity.

\begin{figure}[ht]
\begin{center}  
\includegraphics[scale=0.25, angle=270]{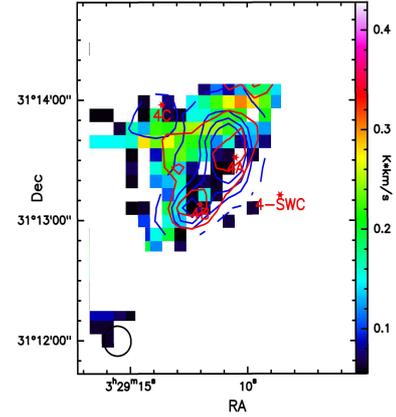}
\end{center}
\caption{Integrated intensity map (from $+$5 to $+$9~km~s$^{-1}$) of H$_{2}$D$^{+}$ 1$_{1,0}$--1$_{1,1}$ in colors, overplotted with N$_{2}$H$^{+}$ 4--3 in blue contours and DCO$^{+}$ 5--4 in red contours. The blue contours are set to 5, 20, 30, 40 $\times$rms (rms: 0.02~K) and the red contours are set to 3, 5 and 8$\times$rms (rms: 0.03~K).}
\label{fig:h2dplus_spa}
\end{figure}

N$_{2}$H$^{+}$ traces very well the distribution of the protostellar envelope and the gas surrounding the protostars, without emission from the IRAS~4--SWC source (Figure~\ref{fig:h2dplus_spa}). The same figure shows that H$_{2}$D$^{+}$ is anti--correlated with the protostar positions, showing emission parallel to the IRAS~4A--4B direction. This covers a part of the area where the DCO$^{+}$ emission is extended. Deuterated species have been found to trace the coldest regions of gas, especially H$_{2}$D$^{+}$ whose main destroyer (CO) is frozen out in the grains \citep{Crapsi05,Caselli1999}. The distribution of H$_{2}$D$^{+}$ points towards a colder layer of gas that could explain the enhanced deuterated species.

The species that were detected in all 3 sources include HCN, HNC, HCO$^{+}$, H$^{13}$CO$^{+}$, DCO$^{+}$, N$_{2}$H$^{+}$, CO, $^{13}$CO, C$^{17}$O, H$_{2}$CO C$_{2}$H, CN and CS. IRAS~4C shows significantly fewer transitions of these species, which are weaker by factors of 2 to 5 compared to IRAS~4A and IRAS~4B (Figure~\ref{fig:speccomp}). The absorption features at 331.2, 358.8, 359.7, 358.9 and 359.6 GHz cannot be identified. They are very broad (FWHM $\sim$75~km~s$^{-1}$), and they are found at the edges of some datacubes, which likely makes them artifacts.

CH$_{3}$OH lines with upper energy levels E$_{up}$$>$35~K are missing from the spectrum of IRAS~4C, while the spectrum of IRAS~4A and IRAS~4B show CH$_{3}$OH lines with E$_{up}$ up to 250~K. This trend is similar for the observed H$_{2}$CO lines towards the 3 sources. IRAS~4C does not show any associated SiO emission. H$_{2}$D$^{+}$ does not really peak toward any of the sources, but its spatial distribution reveals that it covers the space between them (Figure~\ref{fig:h2dplus_spa}). 
A weak emission of 0.08$\pm$0.03~K ($<$3~RMS) is observed towards IRAS~4C but not towards the other sources. Some isotopologues such as H$^{13}$CN, H$_{2}$$^{13}$CO, C$^{34}$S, and deuterated species such as D$_{2}$CO and HDCO are also not detected in the spectrum of IRAS~4C (see \S2 for noise levels), while they are present in the spectrum of IRAS~4A and IRAS~4B. 

The new source, IRAS~4--SWC, is present in multiple observations including our maps of CH$_{3}$OH and H$_{2}$CO for E$_{up}$$<$60~K (RA = 03:29:08.6, Dec = 31:13:12.6 (J2000)). We aim to clarify the nature of the IRAS~4--SWC emission. If it is result of a separate dense core (prestellar or protostellar) we would expect a compact detection in continuum observations. Prestellar cores are typically detected in (sub--)millimeter dust continuum emission, in absorption at mid-- and far--infrared wavelengths \citep{Bacmann2000} and often show evidence of infall motions \citep{Gregersen2000}. A protostar on the other hand, is associated with compact radio/centimeter continuum source accompanied with molecular outflows (i.e. CO), and/or evidence of an internal heating source (e.g. near/mid infrared emission). IRAS~4--SWC is not seen in mid and far--infrared wavelengths as compact emission \citep[e.g. SCUBA, Spitzer;][]{Sandell2001} nor in absorption. Dense cores of 0.3$\msun$ at similar distances in Perseus are found to show a peak flux density of $\sim$100 mJy/beam at 1.1 mm ($\sigma$$\sim$15~mJy/beam) after adopting a temperature of 15~K which is the value we determined for IRAS~4--SWC \citep{Enoch2008}. Given the lack of the predicted emission we conclude that IRAS~4--SWC is not a separate dense core. The fact that IRAS~4--SWC is mainly present only in outflow tracers (e.g. CH$_{3}$OH, SO) suggests that it could be an internal shock at a position of enhanced density in the IRAS~4A outflow.

\vspace{0.3cm} 

\subsection{Line profiles}

\begin{figure}[ht]
\begin{center}
\includegraphics[width=2.5in, angle=0]{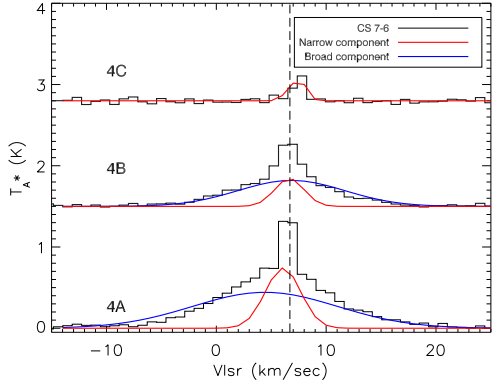}
\end{center}
\caption{Gaussian fit of a broad and a narrow component of CS 7--6 towards IRAS~4A, IRAS~4B and IRAS~4C sources (bottom to top). Dense gas tracers such as CS 7--6 do not show absorption but do show differences in line widths and strengths between the Class 0 objects IRAS~4A, IRAS~4B and the IRAS~4C source (single component).}
\label{fig:cs}
\end{figure}

\begin{figure}[ht]
\begin{center}
\includegraphics[width=2.5in, angle=0]{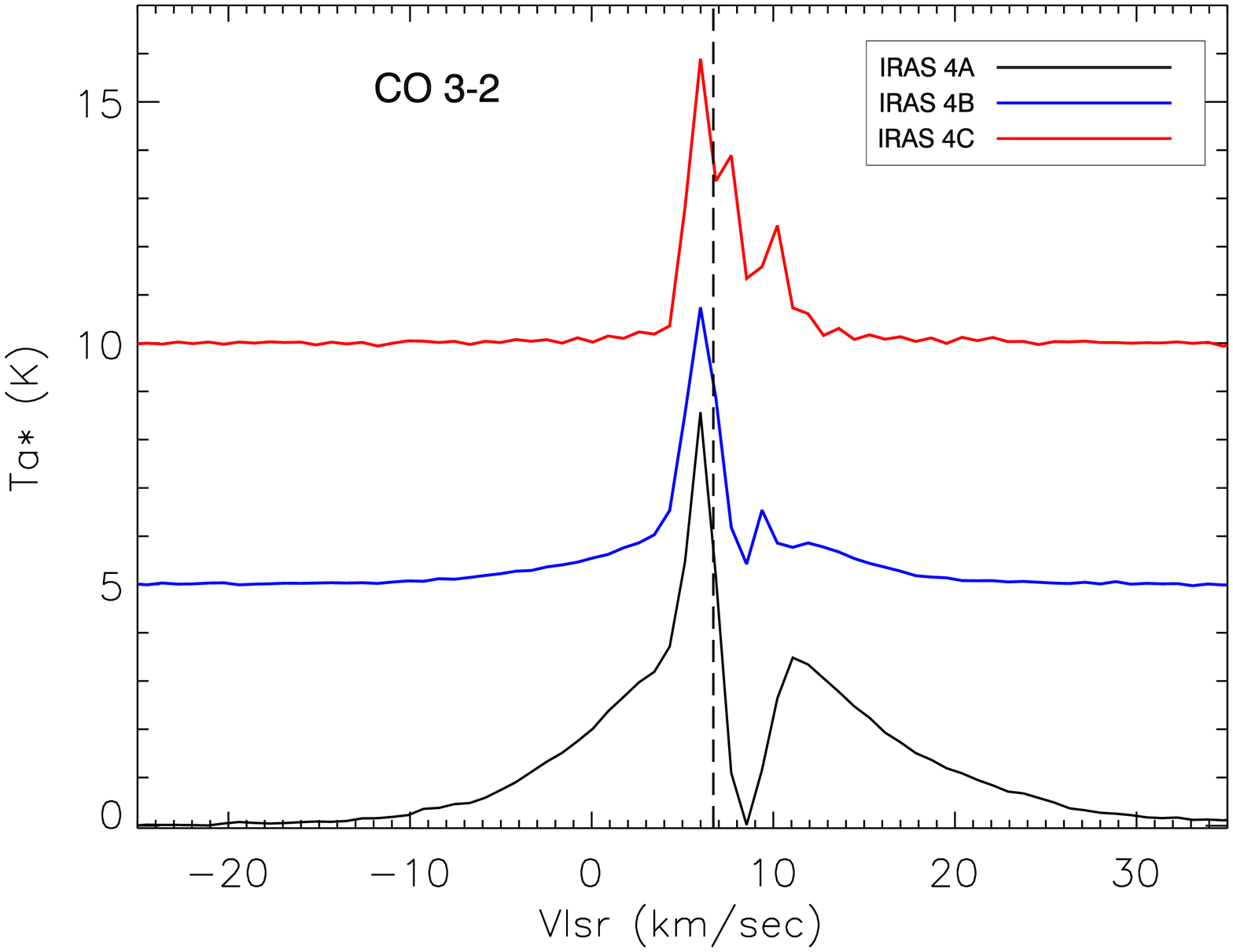} \\
\includegraphics[width=2.5in, angle=0]{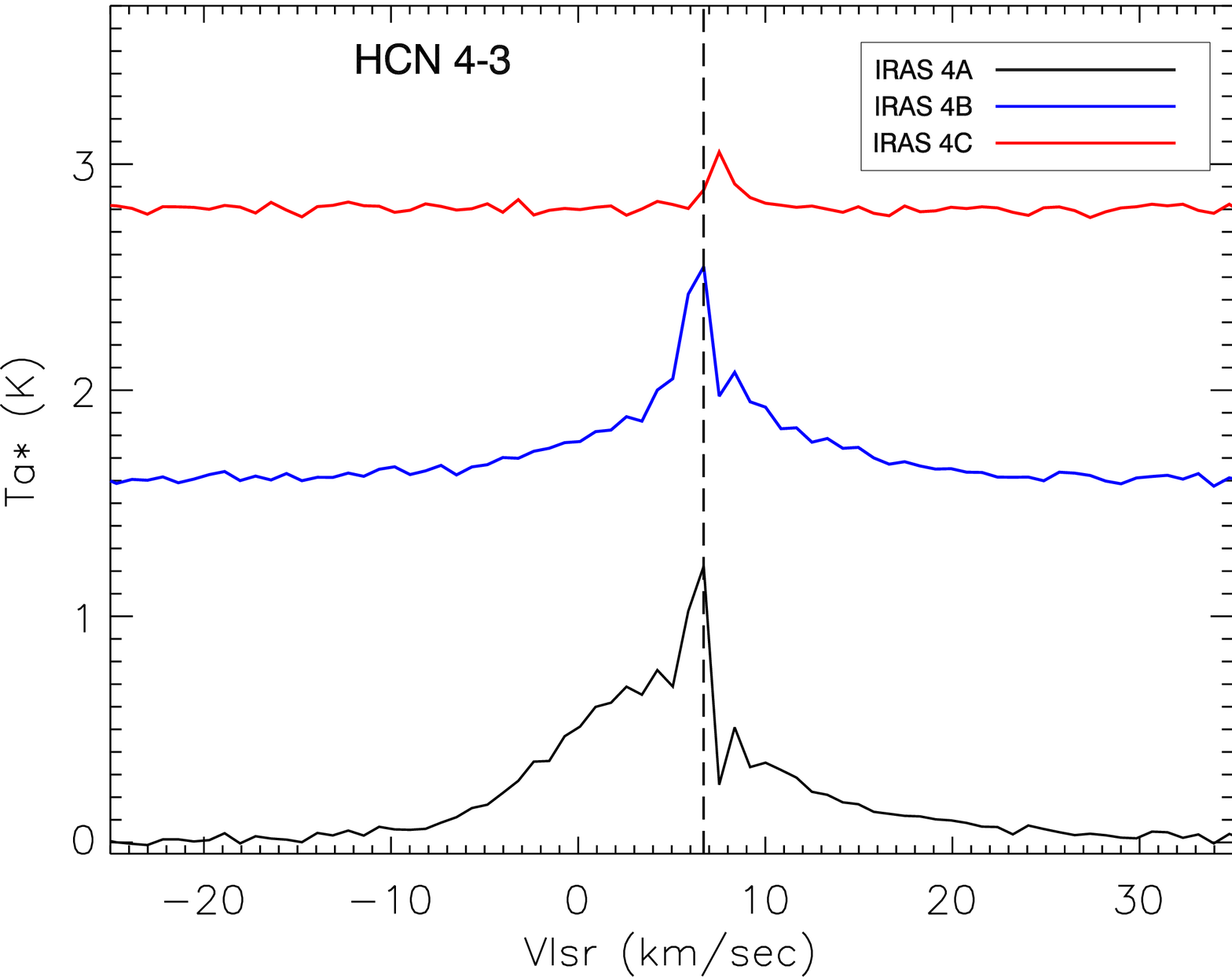} \\
\end{center}
\caption{CO and HCN line profiles for IRAS~4A, IRAS~4B and IRAS~4C sources (bottom to top). Note the absorption towards all positions. IRAS~4A shows a broad wing component, followed by IRAS~4B but IRAS~4C shows a steep line profile and only a very narrow red wing in both CO and HCN. The heavy absorption towards IRAS~4C makes it difficult to fit accurate Gaussians. The vertical line at $+$6.7 km/sec indicates the ambient velocity of IRAS 4A and 4B.}
\label{fig:CO_CS}
\end{figure}

\begin{figure}[ht]
\begin{center}
\includegraphics[width=2.5in, angle=0]{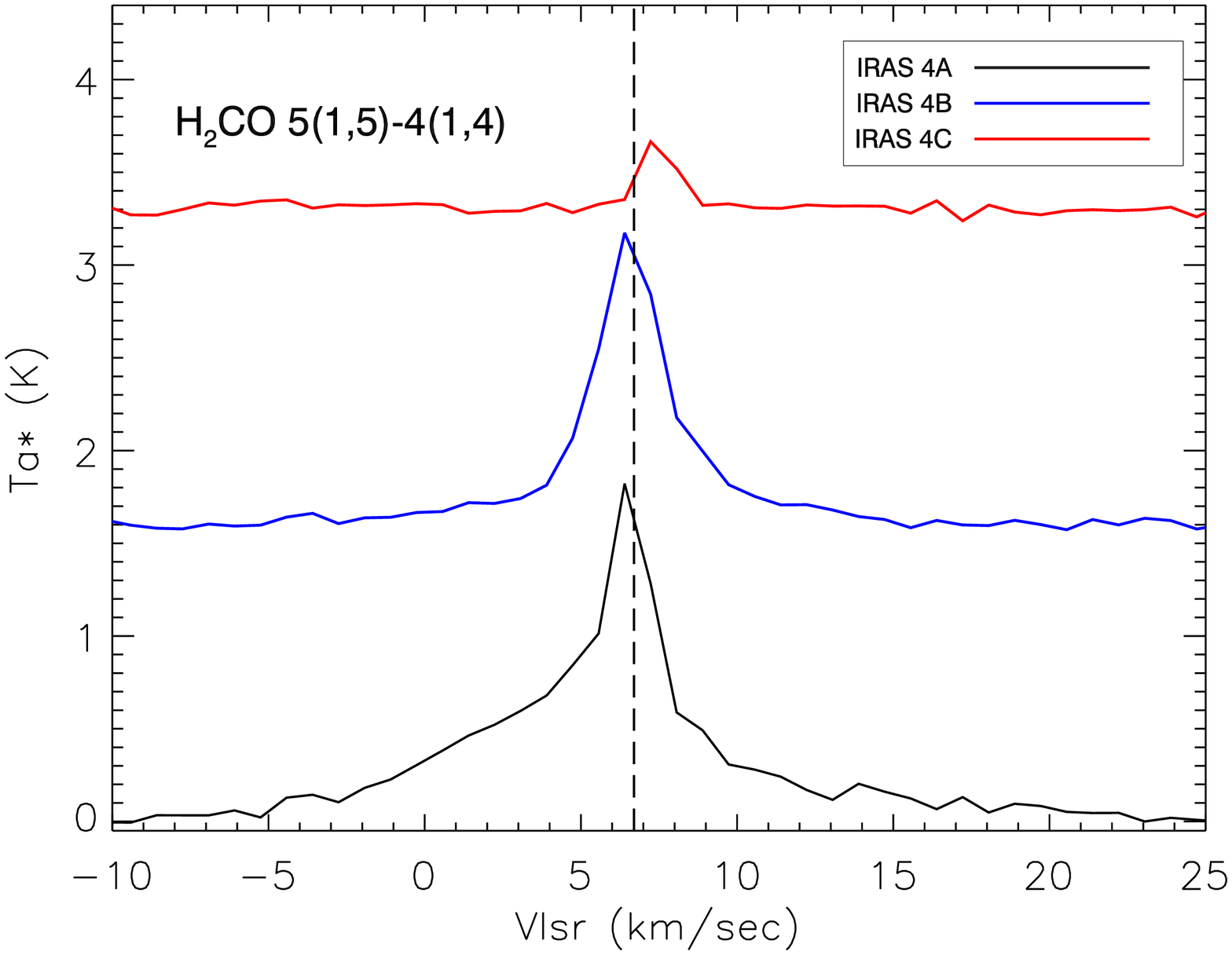} \\
\includegraphics[width=2.5in, angle=0]{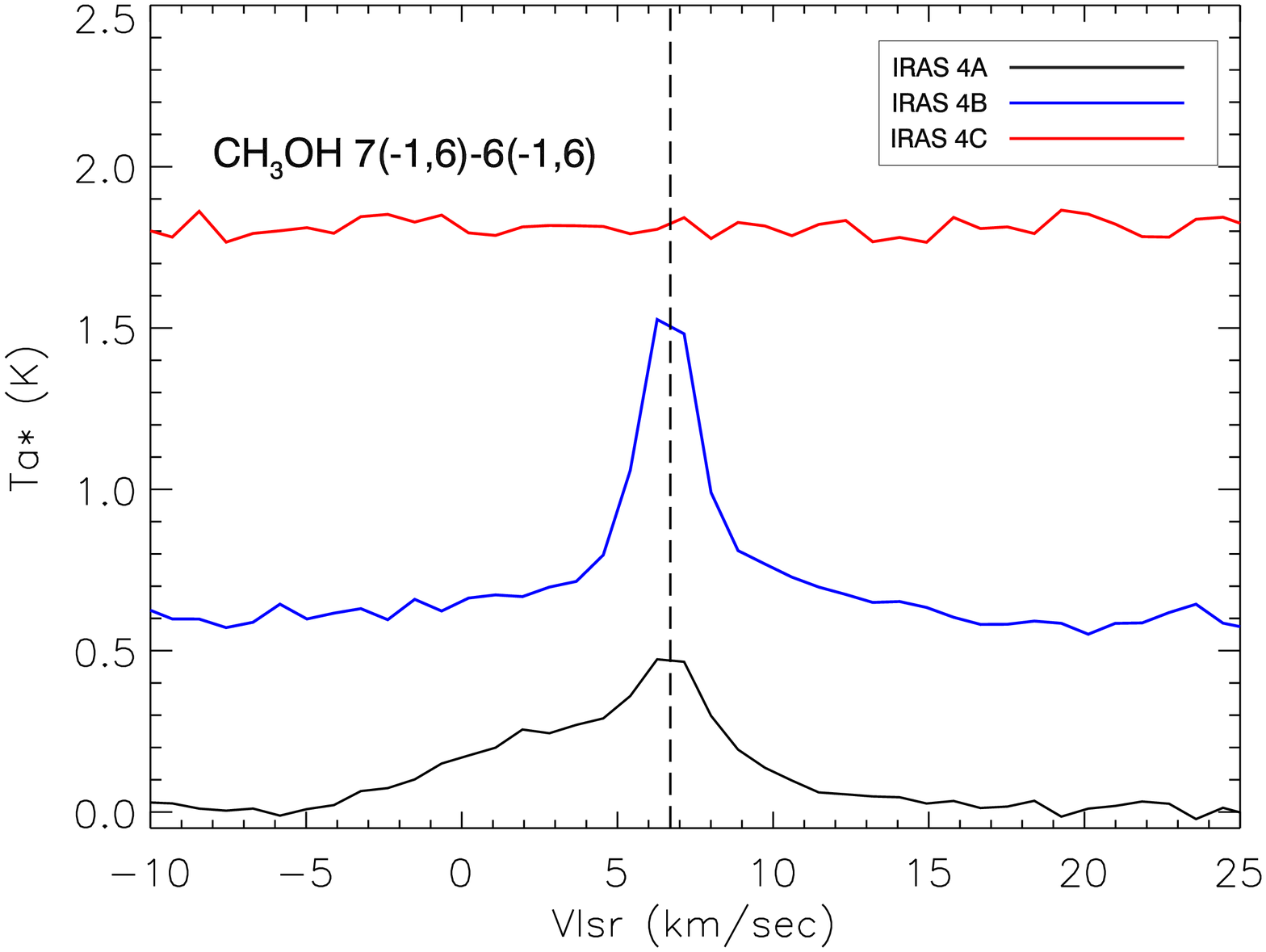}
\end{center}
\caption{Profiles of formaldehyde (top) and methanol (bottom) lines as observed towards the 3 sources IRAS~4A, IRAS~4B and IRAS~4C. IRAS~4C is present at the lower transitions where IRAS~4A is the strongest peak. At higher transitions IRAS~4B is the strongest peak. In all cases IRAS~4A shows the broadest profile among the 3 sources.}
\label{fig:h2co_ch3oh}
\end{figure}

Figures~\ref{fig:cs},~\ref{fig:CO_CS},~\ref{fig:h2co_ch3oh} show representative examples of the observed line profiles towards IRAS~4A, IRAS~4B and IRAS~4C. In most cases, IRAS~4C shows narrow lines that can be fitted with a single Gaussian ($<$2~km~s$^{-1}$) and are 2 to 5 times weaker than the other two sources (Table~\ref{JCMT_lines_1}--\ref{JCMT_lines_3}). There are only few exceptions, e.g. C$^{17}$O, which is 2 times stronger towards IRAS~4C than IRAS~4B and has about the same intensity as towards IRAS~4A. \citet{Shah2001} also reported narrow lines towards IRAS~4C (NH$_{3}$; 1.4~km~s$^{-1}$). 

The majority of the lines towards IRAS~4A and IRAS~4B, can be fitted by two Gaussian components, a narrow (1--3~km~s$^{-1}$) and a broad (5--12~km~s$^{-1}$) with the broadest towards IRAS~4A tracing outflow activity. We observe 5 to 10 times broader lines in IRAS~4A and IRAS~4B compared to IRAS~4C. The fact that the line profiles from IRAS~4C do not show any signature of a broader outflow component (e.g. Figure~\ref{fig:cs}) argues against an outflow along the line of sight. In the alternative case of an outflow in the plane of the sky, one would expect to observe lobes related to the outflow in the CO maps, which is not the case (Section~\ref{sec:maps}). The cone-like structure observed in Figure~\ref{fig:CO_3_2} to the east of IRAS~4C could be a single lobe of an outflow from IRAS~4C, which is the nearest source. If IRAS~4C is at the same distance as IRAS~4A, it is possible that the stronger, more powerful outflow from IRAS~4A would drag the west outflow lobe from IRAS 4C with it, resulting in mixing of the two outflows. This scenario would explain the missing outflow signatures in the line profiles of IRAS 4C for the extreme case that its outflow is located exactly in the plane of the sky, but the fact that the cone does not have its origin exactly in the continuum emission remains puzzling.

The absence of outflow signatures towards IRAS~4C could be a result of its evolutionary status. Broad velocity flows weaken as the core evolves \citep[i.e. from Class 0 to Class I;][]{Arce2007}, which results in less clear spectral line evidence of the flows. In particular the outflow momentum reduces from $\sim$10$^{-2}$~$\msun$~km~s$^{-1}$ to $\sim$10$^{-3}$~$\msun$~km~s$^{-1}$, as we move to more evolved stages \citep{Machida2013}. Outflows are also weaker in very early stages, in particular in the transition from a prestellar core to a Class 0 protostar, resulting in also less broad lines, owing to the lower masses involved in turbulence and/or infall. Thus the absence of outflow signatures towards IRAS~4C could mean either an earlier or a later evolutionary stage of the source compared to IRAS~4A and IRAS~4B. Orientation effects and/or mass may also play a crucial role in the observed differences.

Figure~\ref{fig:CO_CS} presents the line profiles of HCN and CO. 
HCN shows an absorption feature towards IRAS~4A and IRAS~4B but not towards IRAS~4C while CO shows absorption features in all sources. We observe a velocity offset of $\sim$1~km~s$^{-1}$ between the peak intensities of IRAS~4A and IRAS~4B ($+$6.7~km~s$^{-1}$) and IRAS~4C ($+$8~km~s$^{-1}$). The absorption features can be interpreted as infall motions towards IRAS~4A and IRAS~4B which appear as inverse P--cygni profiles. In that case IRAS~4C does not show infall. Another possible explanation for the observed absorption is that there is a foreground gas and IRAS~4C is part of this foreground. The fact that the absorption of both sources appears at similar velocity with the peak intensities of IRAS~4C though, favors the foreground scenario.

Figure~\ref{fig:h2co_ch3oh} shows examples of H$_{2}$CO and CH$_{3}$OH line profiles towards the 3 sources. Striking is the 1.5 to 2.5 times weaker emission of H$_{2}$CO and CH$_{3}$OH transitions toward the IRAS~4C source in comparison with the other 2 sources. We do not detect H$_{2}$CO transitions with E$_{up}$$>$65~K towards IRAS~4C. Regarding methanol (CH$_{3}$OH) lines, IRAS~4B shows the strongest emission among the sources for all observed transitions, while IRAS~4A shows about 2 times weaker emission but 1.5 to 2 times broader lines compared to IRAS~4B. IRAS~4C shows only CH$_{3}$OH transitions with E$_{up}$$<$35~K. 
The methanol lines toward IRAS~4B show a red wing $\sim$7~km~s$^{-1}$ at lower energies (E$_{up}$$<$100~K) indicative of an outflow activity and mostly a single component at higher energies (100~K $<$E$_{up}$ $<$ 200~K), which possibly arises in the dense quiescent parts of the protostellar envelope. The red wing at lower energies and the absence of blue component could be a result of absorption due to a dense, expanding, inner envelope. The methanol lines appear stronger in peak intensity towards IRAS~4B indicating a higher abundance, while in IRAS~4A the wings are broader and show a blue shifted wing indicating faster/stronger outflows. The formaldehyde lines follow the same pattern (Figure~\ref{fig:h2co_ch3oh}).

\begin{figure}[h]
\begin{center}
\includegraphics[width=2.3in, angle=90]{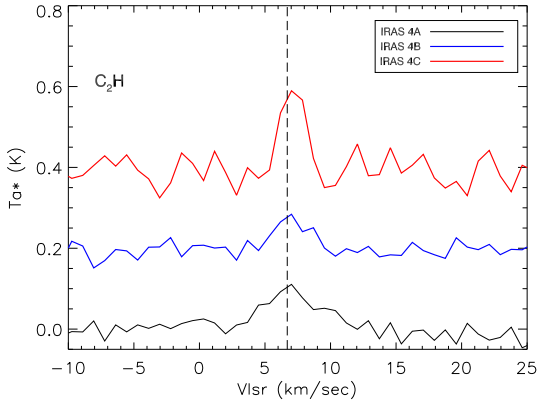}
\end{center}
\caption{Line profiles of C$_{2}$H (N=4-3, J=9/2-7/2, F= 4-3) towards IRAS~4A, IRAS~4B and IRAS~4C (bottom to top). C$_{2}$H is stronger towards IRAS~4C. The intensity of the other 2 lines are comparable, but IRAS~4A shows broader profile. The vertical line represents the ambient velocity of IRAS 4A and 4B at $+$6.7~km~s$^{-1}$.}
\label{fig:c2h_line}
\end{figure}

C$_{2}$H is one of the few species that show the strongest emission toward IRAS~4C (Figure~\ref{fig:c2h_line}). This species is thought to trace early stages of massive star formation \citep{Beuther08}. However, \citet{Sakai2010} report very bright C$_{2}$H towards the low mass late Class~0 source L~1527. More likely the presence or absence of a particular species may signify some different chemistry rather than the final mass of a protostar, while for individual lines, excitation effects may play a role. 

\subsection{Velocity structure}

\label{velo_struc}

\begin{figure}[ht]
\begin{center} 
\includegraphics[width=2.8in, angle=270]{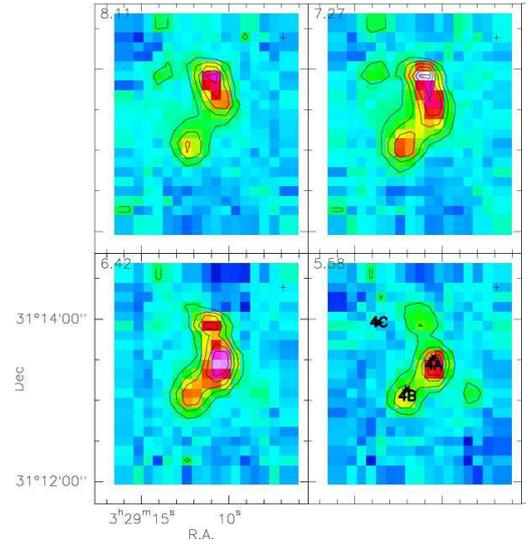}
\end{center}
\caption{Channel maps of SO 9(8)--8(7) obtained with JCMT, with the central velocities given in km$\;$s$^{-1}$. The emission from IRAS~4C appears at higher velocity channels (from $+$7 to $+$8~km~s$^{-1}$) while IRAS~4--SWC at lower velocity channel ($+$5.6~km~s$^{-1}$). The peak north to IRAS~4A at about $+$7~km~s$^{-1}$ can be associated with a outflow/shock from IRAS~4A.}
\label{fig:so}
\end{figure}

The velocity distribution maps of several species (e.g. SO; Figure~\ref{fig:so}) show clearly that although 
the peaks for IRAS~4A and IRAS~4B appear at similar velocity ($+$6.7~km~s$^{-1}$), IRAS~4C and its surrounding area peak at about 1~km~s$^{-1}$ higher velocity ($+$7.9~km~s$^{-1}$). 

IRAS~4--SWC peaks at about 1~km~s$^{-1}$ lower velocity ($+$5.6~km~s$^{-1}$) which indicates an overall velocity gradient in the NE--SW direction ($\Delta$V $\sim$2.5~km~s$^{-1}$), probably due to the structure of the ambient cloud. NGC~1333 is characterized mostly by the $+$8~km~s$^{-1}$ velocity component and it has been suggested that the IRAS~4 core 
($+$6.7~km~s$^{-1}$), is actually a smaller embedded cloud with a different velocity \citep{Choi2004, Langer1996}. The observed HCN absorption towards IRAS~4A and IRAS~4B at $+$7.84~km~s$^{-1}$ (Figure~\ref{fig:CO_CS}) is at a very similar velocity as the peak velocity ($\sim$8~km~s$^{-1}$) of almost all the observed lines towards IRAS~4C (Table~\ref{JCMT_lines_3}). This coincidence could indicate that IRAS~4C might not be a member of the IRAS~4 cloud, but in the foreground. Although the observed differences may lead to question the membership of IRAS~4C in the IRAS~4 cloud, they are small enough to conclude that IRAS~4C is not at a significantly different distance compared to the other two objects.

\subsection{Interferometric continuum and line emission}

\begin{figure}[ht]
\begin{center} 
\includegraphics[scale=0.42]{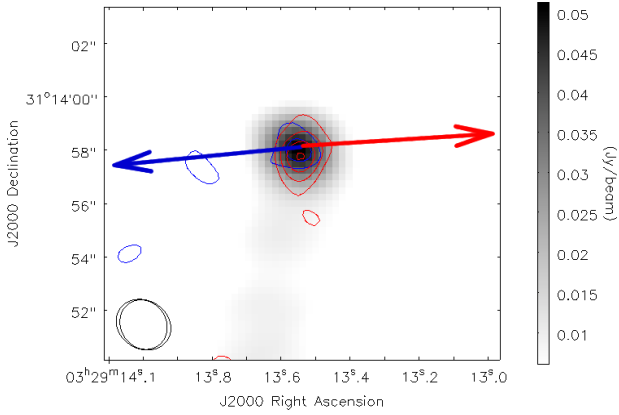}
\end{center}
\caption{CARMA continuum observations of NGC~1333/IRAS~4C at 1.3~mm (grey--scale) overplotted with the red and blue wings of $^{13}$CO 2--1. The velocity range of the line is between $+$5 and $+$11~km~s$^{-1}$, and a siginificant part of the emission has been resolved--out by the interferometer at the source velocity, while we still get emission from the wings. The peaks of the blue (from $+$5 to $+$6.5~km~s$^{-1}$) and red (from $+$8.5 to $+$11~km~s$^{-1}$) wing show a small offset compared to the peak intensity of the continuum. The blue contours levels are set to 0.04 0.06 0.08 0.10 and 0.13 Jy/beam (rms$\sim$0.014~Jy/beam) and the red at 0.06, 0.09, 0.12, 0.15, 0.18~Jy/beam (rms$\sim$0.02~Jy/beam). The ellipse shows the beam size and shape at this wavelength (2.12$\arcsec$$\times$1.81$\arcsec$). The arrows represent the direction of blue--shifted and red--shifted outflow components \citep{Tobin2015}.}
\label{fig:carma_dust}
\end{figure}

Our CARMA continuum observations at 1.3~mm reveal the dust emission from IRAS~4C which appears to trace a compact region (Figure~\ref{fig:carma_dust}). A 2d Gaussian fit of this emission gives a deconvolved component size of FWHM 1.79\arcsec $\times$ 1.35 \arcsec (420$\times$320~ AU). The observed peak flux density of the continuum observations is 51 mJy/beam with a rms of $\sim3$mJy. \citet{Tobin2015} had about 3 times higher resolution resulting in consequently a smaller deconvolved size of FWHM 0.51\arcsec $\times$ 0.22 \arcsec (120$\times$52~AU). \citet{DiFrancesco01} using IRAM PdBI (Beam FWHM: 2\arcsec $\times$ 1.7 \arcsec) towards IRAS~4A and IRAS~4B report a deconvolved size of 920$\times$720~AU and 600$\times$560~AU respectively. The reported differences are a result of the different angular resolutions among the studies.

\begin{figure}[ht]
\begin{center}
\includegraphics[scale=0.32]{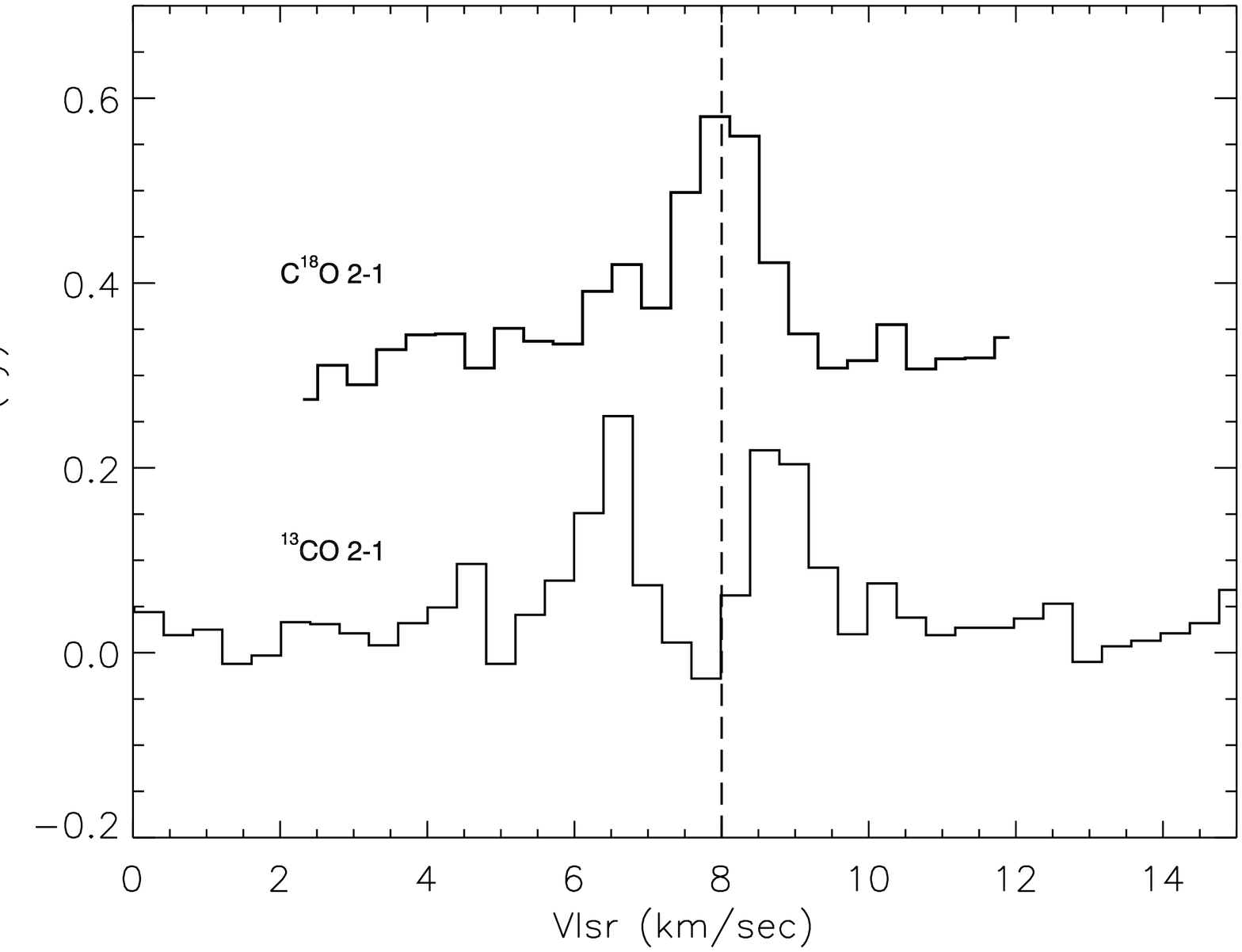}
\end{center}
\caption{CO isotopologues as observed with CARMA towards NGC~1333/IRAS~4C. $^{13}$CO shows broader profile and significant part of the emission has been possibly resolved--out by the interferometer, while C$^{18}$O shows only narrow emission. The vertical line at $+$8~km~s$^{-1}$ is the velocity of the C$^{18}$O peak. The plotted flux offset between the lines has been chosen for easier comparison. $^{12}$CO is not shown as it is heavily affected by absorption making its use very difficult.}
\label{fig:co_carma}
\end{figure}

The 2--1 transitions of CO, $^{13}$CO and C$^{18}$O 
have also been detected in our CARMA data. The integrated intensity peaks of the blue (from $+$5 to $+$6.5~km~s$^{-1}$) and red (from $+$8.5 to $+$11~km~s$^{-1}$) wing of $^{13}$CO show a small offset ($<$1\arcsec) compared to the peak intensity of the continuum (Figure~\ref{fig:carma_dust}), with the blue to be in similar direction with the outflow as indicated by Spitzer observations (further discussion in Sec.~\ref{outflows}). The CO and $^{13}$CO lines show strong absorption at source velocities ($\sim$$+$8~km~s$^{-1}$) possibly as a result of resolved--out emission by the interferometer while C$^{18}$O shows only narrow emission with a FWHM of 1.7$\pm$0.2~km~s$^{-1}$ at $+$7.9$\pm$0.1~km~s$^{-1}$ (Figure~\ref{fig:co_carma}). Similar velocity offsets are observed also with our JCMT data (e.g. Figure~\ref{fig:cs}).\\

\begin{figure}[ht]
\begin{center}
\includegraphics[scale=0.45,trim={0 250 0 150},clip=true]{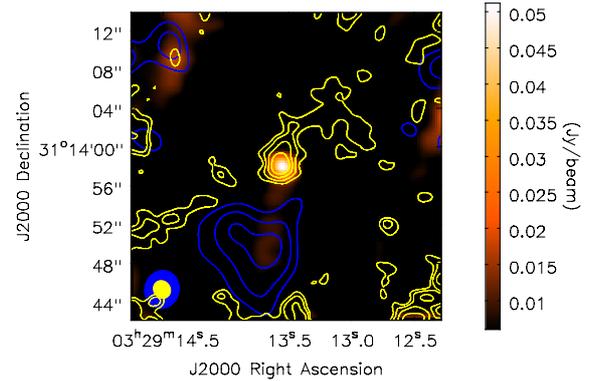}
\end{center}
\caption{Continuum emission towards IRAS~4C in colors, overplotted with the N$_{2}$D$^{+}$ 3--2 emission with blue contours at 0.45, 0.6, 0.7, 0.8 and 0.95 Jy/beam (rms$\sim$15~mJy/beam) and with the C$^{18}$O 2--1 emission with yellow contours at 0.04, 0.08, 0.10, 0.14, 0.20 and 0.29 Jy/beam (rms$\sim$14~mJy/beam). We observe an offset between the continuum source and the N$_{2}$D$^{+}$ of 8.5\arcsec. The beam sizes of the continuum (yellow) and the lines (blue) are also plotted with in the bottom left corner.}
\label{fig:n2dplus_carma}
\end{figure}

We obtain a $\sim$7$\sigma$ detection of N$_{2}$D$^{+}$ near IRAS~4C, which has been previously observed towards all 3 sources \citep[single dish;][]{Friesen2013}. Our maps reveal an offset between the continuum source and the N$_{2}$D$^{+}$ emission of 8.5$\arcsec$, which corresponds to 1900~AU (Figure~\ref{fig:n2dplus_carma}). An offset up to 2000~AU between N$_{2}$H$^{+}$ 1--0, that traces the envelopes of the protostars and the N$_{2}$D$^{+}$ 2--1, 3--2 has been previously reported by \citet{Tobin2013a} towards 8 protostellar objects and by \citet{Lee2015} towards 3 protostars in Perseus. N$_{2}$D$^{+}$ is expected to arise from environments where CO is depleted. Our finding suggests that this happens in a region offset from the dense core.

\section{Evolutionary constraints} 

In our attempt to distinguish the evolutionary status among the 3 sources we mainly study the 
properties of the outflows which are found to have speeds of 10--20 km~s$^{-1}$ \citep{Arce06} for Class~0 objects and maximum of 3~km~s$^{-1}$ for 
a FHSC \citep{Machida2008,Tomida2010}, the temperature among the sources which increases as a protostar evolves, and the deuterium fractionation 
which can be used as a chemical clock \citep{Belloche06,Fontani2011,Fontani14}. 

\subsection{Properties of the outflows}

\label{outflows}

We have mapped the NGC~1333 IRAS~4 region in CO 3--2 (JCMT; Figure~\ref{fig:CO_3_2}). The spectra show deeply self--absorbed line towards all the 3 sources, broad wings towards IRAS~4A (FWHM$\sim$22.0$\pm$2~km~s$^{-1}$) and IRAS~4B (FWHM$\sim$15.0$\pm$1~km~s$^{-1}$) and a weak red wing signature from IRAS~4C (FWHM$\sim$5.0$\pm$0.4 km s$^{-1}$). The accurate fitting with Gaussians is difficult due to the heavy absorption, especially towards IRAS~4C which shows a much narrower line profile (Figure~\ref{fig:CO_CS}). Our $^{13}$CO 3--2 JCMT line profiles though, can be fitted well with a single narrow component of FWHM$\sim$1.9$\pm$0.1~km~s$^{-1}$ towards IRAS~4C, while IRAS~4A can be fitted with a 2--component Gaussian, a narrow one of FWHM$\sim$2.0$\pm$0.1~km~s$^{-1}$ and a broad one of FWHM$\sim$10.0$\pm$2~km~s$^{-1}$. The line profile of IRAS~4B can also be fitted with only a single component of FWHM$\sim$2.4$\pm$0.1~km~s$^{-1}$.


To search for compact outflows which single-dish observations may not pick up, 
high--resolution interferometric CO or H$_{2}$O observations toward protostellar cores are required (CARMA; Figure~\ref{fig:carma_dust}).

$^{13}$CO (J = 2-1) traces the low--velocity outflow in most class 0 protostars \citep{Arce06} but has been observed also in the inner envelope and the disk of extremely young protostars (e.g. L1527 IRS).
To search for an outflow towards IRAS~4C, we use the $^{13}$CO line and integrate the velocities of the emission excluding the lack of emission which is possibly due to the spatial filtering by interferometer from $+$6.5~km~s$^{-1}$ to $+$8.5~km~s$^{-1}$ (Figure~\ref{fig:co_carma}). We get a narrow velocity range from $+$5 to $+$6.5~km~s$^{-1}$ for the blue shifted emission and from $+$8.5 to $+$11~km~s$^{-1}$ for the red shifted emission and we 
observe a small offset compared to the dust continuum emission, indicating a potential low--velocity outflow (Figure~\ref{fig:carma_dust}). We find an offset to the blue and red component of $^{13}$CO that corresponds to a P.A. of $\sim$29$\degree$ while \citet{Tobin2015} found a P.A. of $\sim$-21$\degree$ for C$^{18}$O 2--1 using the B and C--array CARMA configuration. In that work, this offset is interpreted as a potential rotation signature, but due to poor S/N a Keplerian rotation could not be tested. A VLA 8--mm image of IRAS~4C at $\sim$0.08$\arcsec$ resolution shows also a dust emission connected to a disk--structure and not outflow (Segura--Cox et al. 2015, submitted). Spitzer data have revealed nebulosity showing a scattered light cone which has an origin at 
the location of the protostar \citep[e.g. Fig. 19;][]{Tobin2015}. Thus the most likely direction of the outflow is orthogonal to the plane of the disk. This is not the direction that our $^{13}$CO shows and we measure an offset $\sim$50$\degree$ between our $^{13}$CO and the C$^{18}$O which indicates that the 2 isotopologues do not trace exactly the same gas. 

We measure $\Delta$V $=$1.5~km~s$^{-1}$ for the blue shifted emission and $\Delta$V $=$2.5~km~s$^{-1}$ for the red shifted emission. Some amount of the wing emission might also have been resolved--out by the interferometer. Still the observed velocity range for the potential outflow is very narrow indicating at best a very slow outflow ($\sim$2~km~s$^{-1}$). 

The characteristic outflow velocity V$_{\rm obs}$ that we measure provides a lower limit which can be corrected to the real flow velocity V$_{\rm flow}$ 
if the inclination angle $i$ is known, using

\begin{equation}
V_{\rm flow} = \frac{V_{\rm obs}}{\sin i}
\label{inclination}
\end{equation}

Following the suggestion from \citet{Tobin2015} that IRAS~4C contains a disk nearly at edge-on inclination, which corresponds 
to an outflow inclination angle of $<$ 10$^{\circ}$, adopting an angle of 5$^{\circ}$ we derive velocities up to 17~km~s$^{-1}$ for the blue shifted emission and up to 28~km~s$^{-1}$ for the red shifted emission which are an order of magnitude higher than the measured ones. In this case, we can explain the narrow lines but we cannot explain why the outflow is not spatially visible in our larger JCMT maps (Figure~\ref{fig:CO_3_2}). We cannot exclude the scenario that our $^{13}$CO emission is actually tracing the disk and not the outflow as suggested by \citet{Tobin2015}. 

At the same time IRAS~4B shows a compact outflow with an inclination close to 90$^{\circ}$ \citep{Maret2009}, which means that it is almost 
perpendicular to the plane of the sky and the observed velocity is the maximum it can have. The morphology of the IRAS~4A outflow indicates an inclination close to 20$^{\circ}$ which means that the observed velocities are underestimated by about a factor of 3. The inclination of both outflows have been calculated before using the masers associated with them, which provided proper motion and radial velocity measurements \citep{Marvel2008}. In that work both maser outflows were found to be nearly in the plane of the sky (2$^{\circ}$ for IRAS~4A, 13$^{\circ}$ for IRAS~4B) with the IRAS~4B estimate being more uncertain. Although in the studies mentioned above the inclination of IRAS~4B varies a lot, in each case IRAS~4A appears to drive the most powerful outflow close to the plane of sky and thus underestimated. The outflow velocities of IRAS~4A are at least 4 times larger than IRAS~4B which may indicate that IRAS~4A is a less evolved source than IRAS~4B. One must remember that 4A is a binary though. In particular, there is evidence that the source to the NW (A2) has masers while the source to the SE (A1) does not have masers \citep[e.g.][]{Park2007}. The H$_{2}$O masers require the existence of warm dense shocked gas \citep{Elitzur1992}, and mainly disks and outflows in YSOs fulfill these criteria. Thus A2 is probably more evolved than A1 source to the SE.

\subsection{Mean envelope temperature}

\label{h2co}

The determination of the temperature is another crucial parameter that we use to distinguish 
between the different evolutionary stages among objects with the same luminosity. We derive the kinetic temperature of IRAS~4 region using the H$_{2}$CO lines which have been found to be good tracers of kinetic temperatures. We use the most recent collisional rate coefficients as derived by \citet{Wiesenfeld2013} and the Einstein coefficients from the JPL database \citep{Pickett1998}. 


We used the non--LTE radiative transfer program RADEX \citep{vanderTak07} 
to compare the observed integrated intensity fluxes with a grid of models for deriving
kinetic temperatures.
For the model input we used the molecular data from 
the LAMDA database \citep{Schoier05}. 
RADEX predicts line intensities of several molecular transitions for a given set of parameters: 
kinetic temperature, column density, H$_{2}$ density, background temperature, and line width. 

\begin{figure*}
\centering
\includegraphics[width=6cm,angle=180]{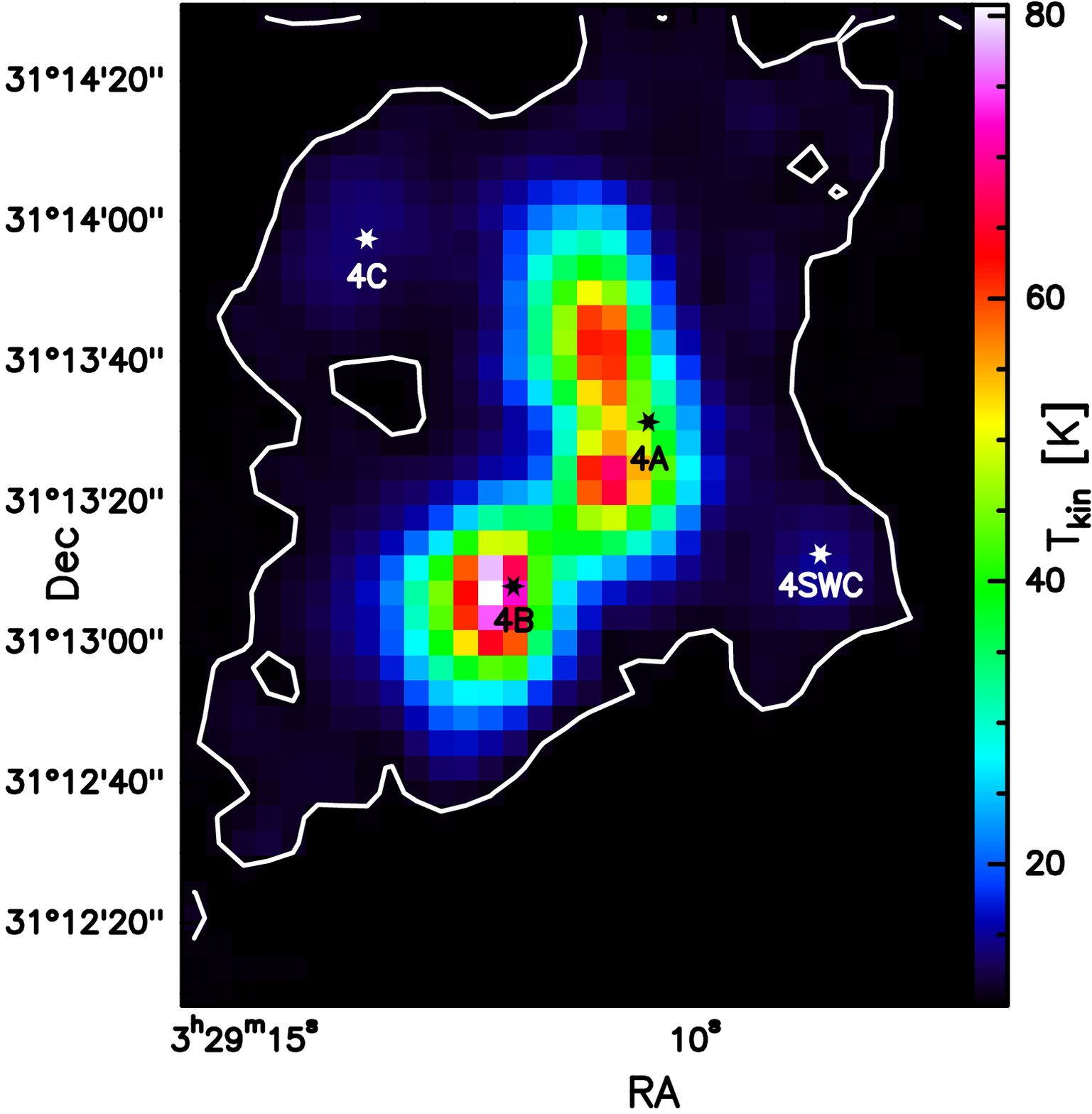}
\includegraphics[width=6cm,angle=180]{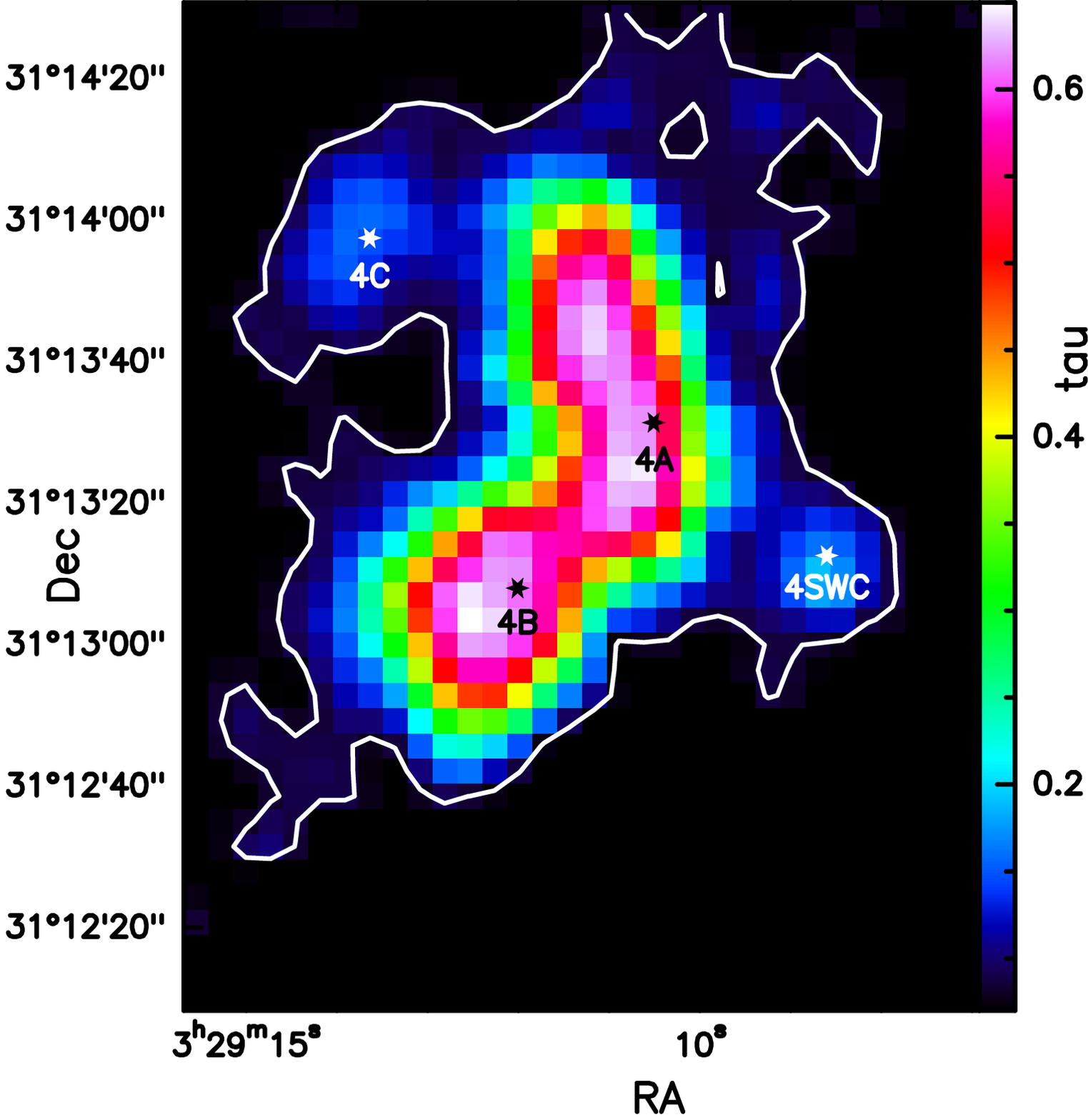}
\includegraphics[width=6cm,angle=180]{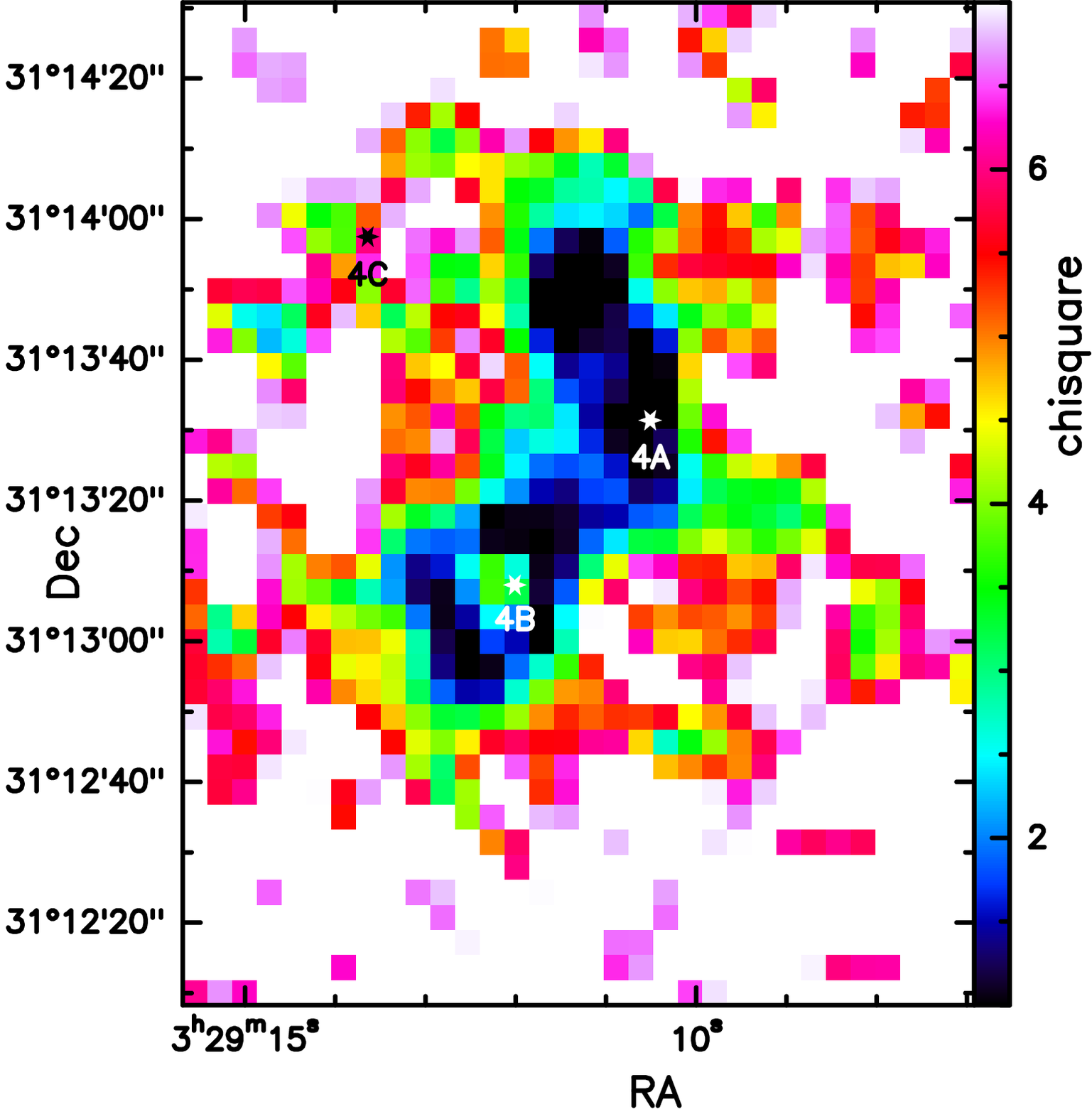}
\caption{Kinetic temperature map of NGC~1333~IRAS~4 (left). During the fitting procedure all the 7 H$_{2}$CO detected lines are taken into account for the positions with S/N $>$3. The white contour shows the region {\bf{where}} at least 2 lines were detected with S/N $>$3. The kinetic temperature is higher towards the Class~0 objects IRAS~4A ($\sim45$~K) and IRAS~4B ($\sim80$~K) while it is significantly lower towards IRAS~4C ($\sim12$~K). Note also the compact emission towards IRAS~4--SWC ($\sim15$~K), which has been observed in channel maps of other molecules (e.g. CH$_{3}$OH, SO, Figure~\ref{fig:so}). The tau map of H$_{2}$CO 5$_{0,5}$--4$_{0,4}$ presents values in the optically thin regime throughout the cloud (middle; $\sim$0.1--0.7). The best resulting $\chi^2$ vary between $\sim$1--5 in the region of the sources (right).}
\label{fig:Tkin}
\end{figure*}

We perform a $\chi^2$ minimization to fit all the observed o--H$_{2}$CO and p--H$_{2}$CO fluxes simultaneously (up to 7 lines, Table~\ref{JCMT_lines_1}), using

\begin{equation}
\chi^2 = \sum\limits_{i=1}^n \frac{(O_{i}-E_{i})^{2}}{E_{i}}
\label{eq:chi}
\end{equation}

integrated over the velocity range from $+$5 to $+$9~km~s$^{-1}$, which corresponds to the narrow emission. In this way we try to limit the contribution of the outflows and perform the analysis under the assumption that we trace the same gas. The $\chi^2$ (Eq.~\ref{eq:chi}) was computed as the quadratic sum of the differences between the observed ($O_{i}$) and the synthetic ($E_{i}$) line intensities for a range of kinetic temperatures (7~K$<$T$_{\mathrm kin}$$<$120~K), a column density of 2$\times$10$^{14}$~cm$^{-2}$ and H$_{2}$ density of 3$\times$10$^{5}$~cm$^{-3}$. These volume and column densities have been obtained from the analysis by \citet{Maret04} towards IRAS~4A and IRAS~4B, which is based on data at similar angular resolution as ours \citep[JCMT, IRAM; 9--17$\arcsec$][]{Maret04}. Adopting these values for the entire cloud suffers from uncertainties (see below) since it is expected that the rest of the cloud is most likely characterized by smaller densities. In addition, the temperature of the background radiation field (CMB) is set to 2.73~K and the line width to 1.8~km~s$^{-1}$ that approximates the value that we have measured throughout the cloud for the narrow component. Finally, in our calculations we assume a fixed ortho to para ratio of o--H$_{2}$CO/p-H$_{2}$CO $=$ 3 and e-CH$_{3}$OH/a-CH$_{3}$OH of 1.

Figure~\ref{fig:Tkin} presents the kinetic temperatures of our modeling results, which vary between $\sim8-80$~K with the highest value stemming from IRAS~4B. The map shows several local peaks with the global one towards IRAS~4B. Core IRAS~4A is not a temperature peak but rather the map shows two peaks northeast ($\sim$15\arcsec) and southeast ($\sim$5\arcsec) of IRAS~4A ($\sim60$~K). 
The gas that surrounds IRAS~4A and IRAS~4B is characterized by T$_{kin}$ $\sim$15--30~K. Two weaker local peaks are observed towards IRAS~4A and IRAS~4--SWC positions. Figure~\ref{fig:Tkin} presents the map of the H$_{2}$CO 5$_{0,5}$--4$_{0,4}$ optical depth, which corresponds to the optically thin regime ($\sim$0.1--0.65) and the best fit $\chi^2$ which is lowest ($\sim$1) towards IRAS~4A and the surrounding gas. Interestingly, the $\tau$ map peaks at the position of IRAS~4A, while T$_{kin}$ map peaks at IRAS~4B. Only IRAS~4C shows up as a local peak in both the optical depth (i.e., column density) and kinetic temperature. The $\tau$ map shows also the two distinctive peaks towards IRAS~4A and IRAS~4B and its entire distribution follows the morphology of the kinetic temperature map. 

Our kinetic temperature estimates towards the Class~0 objects IRAS~4A and IRAS~4B are 45~K and 80~K respectively while it is lower towards IRAS~4C ($\sim12$~K). Deutero--ammonia observations presented in \citet{Shah2001} point also towards very cold conditions towards IRAS~4C (15~K). IRAS~4--SWC is characterized by a kinetic temperature $\sim$15~K and it can be a result of an internal shock connected to the outflow of IRAS~4A or a separate core, as discussed in Sec.~\ref{sec:maps}.

\citet{Maret04} had performed a H$_{2}$CO study towards a sample of 8 low mass protostars, including IRAS~4A and IRAS~4B. 
The collisional rate coefficients that were used in that study were taken from \citet{Green1991}. Our results are in excellent agreement with those 
\citet{Maret04} reported (50~K for IRAS~4A; 80~K for IRAS~4B), especially given the different observations and collision rates used. In their analysis they used pointed observations towards IRAS~4A and IRAS~4B, thus no analysis was performed towards IRAS~4C and their H$_{2}$CO observations were taken from IRAM and JCMT, allowing them to measure more lines but at different angular resolutions. 

To have a sense of the uncertainties in the above calculations, we have run additional models towards IRAS~4A and IRAS~4B, 
where the largest number of lines has been detected. For these models we varied the column density between 2$\times$10$^{12}$~cm$^{-2}$ and 2$\times$10$^{16}$~cm$^{-2}$. We find a best fit column density of 5$\times$10$^{14}$~cm$^{-2}$ for IRAS~4A and 3$\times$10$^{14}$~cm$^{-2}$ for IRAS~4B which are 2.5 and 1.5 times higher than the previously adopted value of 2$\times$10$^{14}$~cm$^{-2}$, but the same order of magnitude. These values result in temperatures of 30~K for IRAS~4A and 64~K for IRAS~4B which are 1.5 and 1.2 times lower than the values we obtained above. IRAS~4B remains warmer than IRAS~4A while their absolute temperature difference remains to be $\sim$35~K, making more prominent the temperature difference among the 2 systems. 

Since we do not have a way to accurately measure the column density of IRAS~4C, and the other 2 sources were found to have higher column density than the adopted, we ran RADEX towards IRAS~4C for a higher N(H$_{2}$CO) adopting a value of 1.5 times higher that before, 3$\times$10$^{14}$~cm$^{-2}$. The increase in column density results in a kinetic temperature of $\sim$8~K which is lower by 4~K compared to our previous estimation and maintains the IRAS~4C the colder among the sources. We also ran our calculations for 1.5 times lower column density using 7.5$\times$10$^{13}$~cm$^{-2}$ a value that results a temperature of $\sim$18~K which is 1.7 times higher than what we obtained in our analysis. Even in that case IRAS~4C remains to be the coldest among the sources with absolute differences of $\sim$20~K and $\sim$40~K with IRAS~4A and IRAS~4B respectively. 

To test the assumption of constant volume density in our analysis, we have re--run our calculations 
for n$_{H_{2}}$ $=$ 10$^{6}$~cm$^{-3}$ and 10$^{4}$~cm$^{-3}$ for 
the entire cloud. The lower value of 10$^{4}$~cm$^{-3}$ was adopted because the cloud around the compact objects is most likely 
characterized by smaller densities. In both cases the distribution of the kinetic temperature is the same as seen in 
Figure~\ref{fig:Tkin}, but the values vary. For the higher density the temperature is characterized by significantly lower values (7~K$<$T$_{\mathrm kin}$$<$22~K). Once more IRAS~4B is the warmest ($\sim$20~K), followed by IRAS~4A ($\sim$16~K) and IRAS~4C ($\sim$8~K). For the lower density the cloud is characterized by temperatures 20~K$<$T$_{\mathrm kin}$$<$82~K resulting $\sim60$~K for IRAS~4B, $\sim35$~K for IRAS~4A and $\sim16$~K for IRAS~4C. This result indicates that our solution is more sensitive to higher densities leading to an overestimation of temperatures for IRAS~4A and IRAS~4C at the adopted n$_{H_{2}}$ $=$ 3$\times$10$^{5}$~cm$^{-3}$, since they were found to be denser in our approach as described in Sec.~\ref{mass}. Although the absolute values for kinetic temperatures vary with the different assumptions, IRAS~4B remains the warmest and IRAS~4C the coldest among the 3 sources in all our calculations.

\subsection{Mass and density}

\label{mass}

We use our CARMA continuum observations to estimate the gas mass (Equation~\ref{mass_gas_1}) and the H$_{2}$ column (Equation~\ref{column_density}) and volume density of IRAS~4C. We calculate these parameters via:

\begin{equation}
M_{gas} = \frac{S_{\nu}d^{2}\alpha}{\kappa_{\nu}B_{\nu}(T_{d})}
\label{mass_gas_1}
\end{equation}


\noindent where S$_{\nu}$ $=$ 0.061 Jy is the total flux density after fitting a 2D--Gaussian of 2.3\arcsec$\times$2.1\arcsec, $\alpha$ $=$ 100 is the gas to dust ratio, $\kappa_{\nu}$(1.3~mm) $=$ 0.89 cm$^{2}$~g$^{-1}$ is the dust opacity per unit mass \citep{Ossenkopf94} and B$_{\nu}$(T$_{d}$) is the Planck function at dust temperature T$_{d}$ ($\sim$12~K; H$_{2}$CO analysis described at~\ref{h2co}). We find a M$_{gas}$ $=$ 0.16 $M_{\odot}$. We adopt a distance $d$ of 235~pc. We calculate the column density using

\begin{equation}
N(H_{2}) = \frac{I_{\nu}\alpha}{2m_{H}\Omega_{b}\kappa_{\nu}B_{\nu}(T_{d})}
\label{column_density}
\end{equation}

\noindent where I$_{\nu}$ is the peak flux density (0.048 Jy/beam), m$_{H}$ is the mass of hydrogen and 
$\Omega$$_{b}$ is the beam solid angle. The resulting N(H$_{2}$) $=$ 1.5$\times$10$^{24}$ cm$^{-2}$ while the volume density assuming 2$\arcsec$ diameter for the adopted area as determined by a 2d Gaussian fit, is$\sim$3.1$\times$10$^{8}$ cm$^{-3}$. This value is higher by 3 orders of magnitude compared to the values obtained by \citet{Maret04} for IRAS~4A and IRAS~4B (3$\times$10$^{5}$ cm$^{-3}$) and the one we have adopted for IRAS~4C for our H$_{2}$CO analysis (3$\times$10$^{5}$ cm$^{-3}$;~\ref{h2co}) under the assumption that the 3 sources are characterized by similar volume densities, but the scales are different since the CARMA beam is $\sim$10$\times$ smaller and thus trace denser region. 

\citet{Smith2000} derived masses of 10.9 $M_{\odot}$ for IRAS~4A, 6.9 $M_{\odot}$ for IRAS~4B and 2.9 $M_{\odot}$ for IRAS~4C using the SCUBA 850$\mu$m continuum map of NGC~1333/IRAS 4, with a FWHM of the beam $\sim$16$\arcsec$ and assuming the same temperature (30~K) for the 3 sources. In that work they had considered the CO 3--2 line contamination in the 850~$\mu$m SCUBA passband, as a possible cause of an underestimation of the calculated spectral index and overestimation of the masses, but it could still not explain the low spectral indices observed towards the three sources. 

The mass of IRAS~4C that we obtain is about an order of magnitude lower but the CARMA beam is $\sim$10 times smaller. In addition we use the most recent distance estimate of 235~pc \citep{Hirota2008}, compared to their distance of 350~pc \citep{Herbig1983}. Correcting for the distance and the temperature and using their reported values for total and peak flux density and their angular resolution we find a volume density of 6$\times$10$^{6}$~cm$^{-3}$, which is about 2 orders of magnitude lower than the one obtained from the CARMA data, providing an evidence of a density gradient at IRAS~4C. We follow the same process for IRAS~4A and IRAS~4B using a FWHM~$\sim$25$\arcsec$ as a result of the Gaussian fit, and we measure the values of 2$\times$10$^{6}$~cm$^{-3}$ for IRAS~4A and 5$\times$10$^{5}$~cm$^{-3}$ for IRAS~4B. So we find that IRAS~4C is 3 times denser than IRAS~4A and IRAS~4A is 4 times denser than IRAS~4B. Since our CARMA observations do not cover IRAS~4A and IRAS~4B we cannot further investigate possible density gradients. 

Lastly, for a direct comparison, we determine the masses of the sources in a resolution similar to our JCMT maps. For this purpose, we use the SCUBA data \citep{Smith2000}, adopting the more recent distance of 235~pc and the temperatures we derived in Sec.~\ref{h2co} and we calculate masses (Eq.~\ref{mass_gas_1}) of 2~$M_{\odot}$ for IRAS~4A, 0.6~$M_{\odot}$ for IRAS~4B and 1.5~$M_{\odot}$ for IRAS~4C. This is the first time that IRAS~4C appears with a higher mass than IRAS~4B due to the fact that we do not use the same temperatures for all 3 sources, as previous studies, but the ones determined in Sec.~\ref{h2co}. The derived masses and column densities are very sensitive to the adopted opacity and temperatures which can cause a difference up to a factor of 3--5 given the temperature uncertainties discussed above.

\subsection{Depletion of CO}

\label{depletion}

Carbon monoxide and its isotopologues are broadly used as a tracer of
N(H$_{2}$) in studies of the interstellar medium. The most
abundant $^{12}$CO, can only provide a lower limit of the column density and the mass of the region since it is often 
optically thick in typical conditions 
of the molecular clouds. The less abundant
isotopologues can then been used (e.g. $^{13}$CO, C$^{18}$O), assuming optically thin emission. 

\begin{figure}[h]
\begin{center}
\includegraphics[scale=0.45, angle=180]{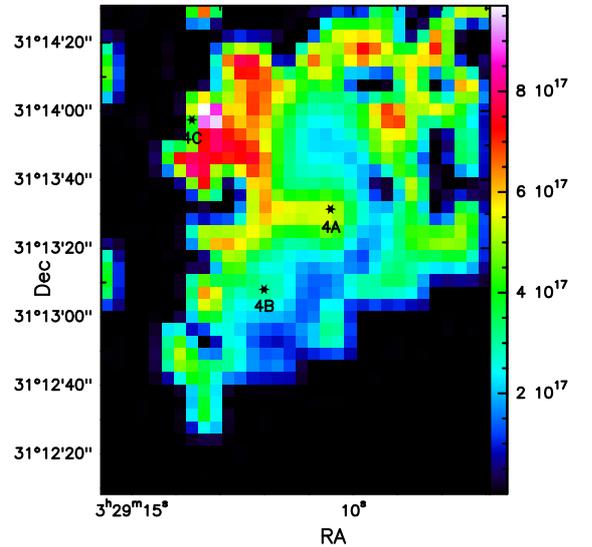} 
\end{center}
\caption{Synthetic CO column density as determined fitting the peak intensities for the range from $+$5 to $+$9~km~s$^{-1}$.}
\label{fig:CO_synthetic}
\end{figure}

To estimate the spatial distribution of N(CO), we use the C$^{17}$O 3-2 line, assuming a constant $^{16}$O/$^{17}$O of 2000 \citep{Wilson1999}. We run RADEX in order to derive the column density of C$^{17}$O, using the gas temperature map as determined from the H$_{2}$CO analysis and a constant volume density of 3$\times$10$^{5}$~cm$^{-3}$. We derive N(CO) for the region where T$_{peak}$$>$3~rms, to be in the range of 2$\times$10$^{17}$~cm$^{-2}$ -- 9$\times$10$^{17}$~cm$^{-2}$ (Figure~\ref{fig:CO_synthetic}). 

\begin{figure}[ht]
\begin{center}
\includegraphics[width=2.5in]{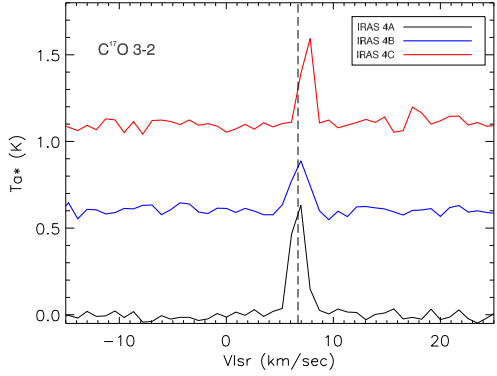}
\end{center}
\caption{C$^{17}$O 3-2 line profiles towards IRAS~4A, IRAS~4B and IRAS~4C. The line is strongest towards IRAS~4A followed by IRAS~4C and then IRAS~4B.}
\label{fig:c17o_line}
\end{figure}

The higher value is close to IRAS~4C. This is a result of low kinetic temperature of IRAS~4C compared to the other two sources ($\sim$12~K) while the intensity of C$^{17}$O line is stronger towards IRAS~4C than towards the warmest IRAS~4B (Figure~\ref{fig:c17o_line}). In order to be able to fit the relatively high intensity towards IRAS~4C, for so low kinetic temperature RADEX requires higher column density, given the fact that we use a constant volume density. 

In order to derive the N(CO)/N(H$_{2}$) for the 3 sources we use Equation~\ref{column_density} applying the peak flux densities reported by \citet{Smith2000} from the 850$\mu$m continuum. In this way we ensure that the angular resolution between continuum and CO observations is about the same (HPBW $\sim$ 13.5 $\arcsec$) and thus that we trace the same amount of gas. For the temperature we use 45~K, 80~K and 12~K for IRAS~4A, IRAS~4B and IRAS~4C respectively as have been determined in Sec.~\ref{h2co}. We find the column density N(H$_{2}$) towards IRAS~4A, IRAS~4B and IRAS~4C to be 2.4$\times$10$^{23}$~cm$^{-2}$, 1.3$\times$10$^{23}$~cm$^{-2}$ and 3.6$\times$10$^{23}$~cm$^{-2}$ resulting N(CO)/N(H$_{2}$) of 2.4$\times$10$^{-6}$, 3.8$\times$10$^{-6}$ and 1.2$\times$10$^{-6}$. These values are $\sim1.5-2$ orders of magnitude less than the canonical [CO]/[H$_{2}$] of 10$^{-4}$ indicating significant depletion of CO. 

The CO depletion factor, f$_{D}$ can be calculated via:

\begin{equation}
f_{D} = \frac{X^{E}_{CO}}{X^{O}_{CO}}
\label{mass_gas}
\end{equation}

\hspace{-0.6cm} where X$^{E}$$_{CO}$ is the expected abundance of CO relative to H$_{2}$ (10$^{-4}$) and X$^{O}$$_{CO}$ is the ratio between the observed N(CO) and the observed N(H$_{2}$). 

We determine a f$_{D}$ of 42, 26 and 83 for IRAS~4A, IRAS~4B and IRAS~4C respectively. The highest degree of CO depletion is observed towards the coldest of the sources, IRAS~4C, while the lowest degree of CO depletion towards the warmest source, IRAS~4B, as expected \citep[e.g.;][]{Bacmann2002}. Our solution is dependent on the assumed parameters, including the resulting kinetic temperature for the 3 sources and the assumption that H$_{2}$CO and C$^{17}$O trace the same gas. As discussed also in Sec.~\ref{h2co}, IRAS~4C is the coldest of the sources under each assumption and thus although the absolute values of depletion factors may vary, the estimated trend between the sources will remain.

\subsection{Deuteration}
\label{sec:Deuteration}

The deuterium fractionation (e.g. [N$_{2}$D$^{+}$]/[N$_{2}$H$^{+}$]) is another factor to estimate the evolutionary stage of IRAS~4C. More specifically, 
the [N$_{2}$D$^{+}$]/[N$_{2}$H$^{+}$] can be used as a chemical clock \citep{Belloche06,Pagani2013,Fontani14} and has been measured to be $\sim 15-20$ \% for prestellar cores 
(e.g. L1544) and $\sim 5-10$ \% for Class 0 objects (e.g. L1521F). A measurement between these two limits is an indication for an object for which the 
evolutionary stage lies between these two early star formation stages. The cosmic ratio [D]/[H]$=$1.5$\times$10$^{-5}$ \citep{Linsky95} and that higher observed values are typical for low--mass dense cores \citep[e.g.;][]{Crapsi05,Parise2006}. This can be explained due to the fact that at very low temperatures the reaction H$_{3}$$^{+}$ $+$ HD $\Longleftrightarrow$ H$_{2}$D$^{+}$ $+$ H$_{2}$ $+$ $\Delta$E is driven strongly to the right increasing the abundance of H$_{2}$D$^{+}$, in combination with the undergoing freeze out of CO which decreases the H$_{2}$D$^{+}$ destruction rate \citep[e.g. prestellar cores][]{Bacmann2003}. This process leads to enhancement of H$_{2}$D$^{+}$ and the D atom passes down to other species leading to formation of more deuterated species such as N$_{2}$D$^{+}$. We do not have N$_{2}$D$^{+}$ observations towards all the sources and our N$_{2}$H$^{+}$ emission is from JCMT data making the determination of [N$_{2}$D$^{+}$]/[N$_{2}$H$^{+}$] impossible for our dataset. 

We used RADEX to model the observed emission in HCO$^{+}$, H$^{13}$CO$^{+}$ and DCO$^{+}$. The collision rates of HCO$^{+}$ and DCO$^{+}$ with H$_{2}$ were adopted from \citet{Flower1999}. The critical densities n$_{cr}$ of HCO$^{+}$ 4--3 and DCO$^{+}$ 5--4 at 50~K are $\sim$9.1$\times$10$^{6}$~cm$^{-3}$ and $\sim$9.3$\times$10$^{6}$~cm$^{-3}$, calculations based on \citet{Schoier05}. 

\begin{figure}[h]
\includegraphics[scale=0.45, angle=180]{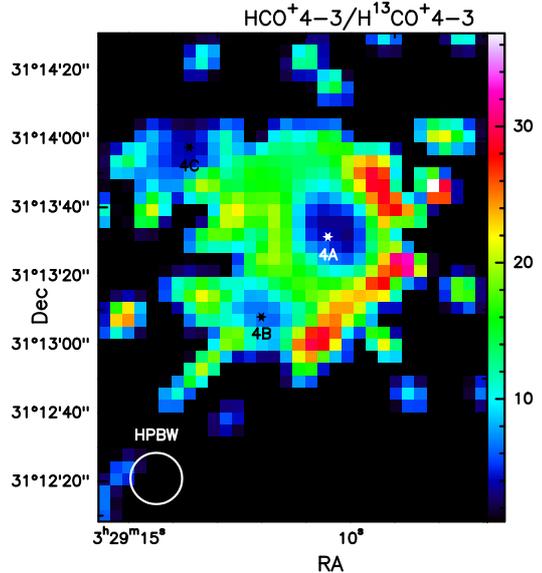} 
\caption{Observed HCO$^{+}$ 4-3 / H$^{13}$CO$^{+}$ 4-3 ratio of the peak intensities for the range from $+$5 to $+$9~km~s$^{-1}$. The fact that 
the observed ratio takes values between 6 and 37, which is smaller than the optically thin ratio of 60, indicates that the main 
isotope HCO$^{+}$ 4-3 is optically thick throughout the cloud. The effect is stronger towards the 3 embedded objects.}
\label{fig:ratioHCOplus}
\end{figure}

For this purpose we used the kinetic temperature map resulting from the H$_{2}$CO analysis and a constant volume density of 3$\times$10$^{5}$~cm$^{-3}$ for the entire cloud. 
Fitting the integrated intensities of the lines for each spatial point enabled us to produce the column density maps 
of H$^{13}$CO$^{+}$, DCO$^{+}$. We find column densities between 8$\times$10$^{12}$--4$\times$10$^{13}$cm$^{-2}$ for H$^{13}$CO$^{+}$ with the higher value close to IRAS~4C and 8$\times$10$^{13}$--1.5$\times$10$^{14}$cm$^{-2}$ for DCO$^{+}$. HCO$^{+}$ is usually optically thick, something that we were able to test by determining the observed 
HCO$^{+}$ 4-3 / H$^{13}$CO$^{+}$ 4-3 ratio of the peak intensities for the range from $+$5 to $+$9~km~s$^{-1}$ (Figure~\ref{fig:ratioHCOplus}). 

The observed ratio is found 
to vary between 6 and 37, which is 10 to 1.5 times smaller than the expected ratio of 60 under optically thin conditions, indicating that the main 
isotope HCO$^{+}$ 4-3 is moderately optically thick throughout the cloud. The effect is stronger towards the 3 embedded objects as expected. For this reason we produce 
the [DCO$^{+}$] / [HCO$^{+}$] towards the 3 sources using a fixed isotopic ratio of [HCO$^{+}$]/[H$^{13}$CO$^{+}$] = 60. 
We reran our calculations adopting the volume densities for each source as determined in Sec.~\ref{mass} and we determine deuteration of 
$\sim$12$\pm2$$\%$ towards IRAS~4A, $\sim$3.5$\pm1$$\%$ towards IRAS~4B and an upper limit of $\sim$20$\%$ towards IRAS~4C. The lowest deuteration is towards IRAS~4B and is correlated also with the lower degree of depletion towards that source as we determine in Sec.~\ref{depletion}. For IRAS~4C the error is an order of magnitude higher compared to the other 2 sources, since at such low temperatures a decrease or increase of temperature by only 5~K causes almost one order and half order of magnitude higher or lower N(DCO$^{+}$) and N(HCO$^{+}$) respectively. One would expect that IRAS~4C would show the higher degree of deuteration since it is the coldest among the sources and thus the enhancement of H$_{2}$D$^{+}$ abundance contributes to the enhancement of DCO$^{+}$ through the reaction H$_{2}$D$^{+}$ $+$ CO $\Longrightarrow$ DCO$^{+}$ $+$ H$_{2}$. Although we report the highest deuteration towards IRAS~4C, this value is only an upper limit and we cannot draw safe conclusions regarding its evolutionary status compared to the other two sources only based on their deuteration.

\subsection{HCN/HNC}

We want to test our temperature estimates by using a method not connected to the H$_{2}$CO lines. 
The [HCN]/[HNC] ratio has been found to be sensitive to temperature \citep{Schilke1992}, while for cold molecular clouds 
is expected to be $\sim$1 \citep{Sarrasin2010}. More precisely HNC/HCN have been found to be decreasing while temperature increases and vice versa \citep{Graninger2014}. 

We follow similar procedure as in Sec.~\ref{sec:Deuteration} using the collisional rate coefficients for HCN and HNC from \citet{Dumouchel2010}, scaled by a factor 1.37 to represent collisions with H$_{2}$. We use the optically thin 
H$^{13}$CN and adopting a HCN/H$^{13}$CN $=$ 60 we present the observed HCN/HNC ratio towards the 3 sources. We find a ratio of $\sim$3.5 for IRAS~4A and $\sim$4.2 for IRAS~4B. The isotopologue HN$^{13}$C did not show a clear detection and thus we assume that HNC is optically thin towards all 3 positions. None of the rarer isotopologues are observed towards IRAS~4C so we assume that the main isotopologues are optically thin and we find there the lowest observed ratio of $\sim$0.85. 

Our results show nicely the correlation between HCN/HNC and the temperature, since the lowest value corresponds to the colder source (4C) and the highest to the warmer source (4B). We also find more HNC than HCN towards IRAS~4C, where the temperature is as low as $\sim$12~K and HNC emission is expected to be higher than HCN for such low temperatures \citep{Padovani2011,Graninger2014}.

\section{Discussion and Conclusions}

IRAS~4C, has been reported based on continuum observations in various previous works as a Class~0 object \citep[e.g.;][]{Enoch09}. We take the advantage of the direct comparison with the other 2 
nearby Class~0 objects, IRAS~4A and IRAS~4B, using single dish mapping observations as they have been obtained during the JCMT line survey. In addition we report the mass and the 
volume density as was determined using our CARMA observations.

\subsection{Results}

We tried to distinguish the evolutionary stage of IRAS~4C compared to the other two Class~0 objects, taking into account the derived physical parameters from our analysis sections, the projection effects, the mass and the luminosity of the sources. 
\begin{itemize}
\item We do not observe extended bipolar emission from the outflow towards IRAS~4C, as we would expect from an edge--on disk structure. IRAS~4A drives the faster outflow, followed by IRAS~4B and IRAS~4C. Outflows could be used as evolutionary constraints in the earlier phases of star formation. 
\item We determine the lowest kinetic temperature towards IRAS~4C ($\sim$12~K), followed by IRAS~4A ($\sim$45~K) and IRAS~4B ($\sim$80~K).
\item We take into account the different derived temperatures among the sources and we find IRAS~4C to have 3 times higher mass than IRAS~4B. This can explain why the line profile of C$^{17}$O is stronger towards IRAS~4C than IRAS~4B. IRAS~4A remains the most massive among the sources (2~$M_{\odot}$). 
\item The warmest source IRAS~4B is characterized by the lowest degree of depletion and deuteration, and the highest HCN/HNC confirming its warm nature. In addition IRAS~4C is characterized by the highest degree of depletion and HCN/HNC $<$ 1 characteristic of objects with very low temperature. Due to large errors at such low temperatures, we can only provide an upper limit for the deuteration towards IRAS~4C ($\sim$20$\%$).    

 \item IRAS~4A is $\sim$2 times less warm, $\sim$3 times more massive with $\sim$4 times stronger outflow activity and $\sim$3 times lower deuteration than IRAS~4B, and it is probably younger than IRAS~4B. 
\item Our velocity distribution maps of several species show that IRAS~4A and IRAS~4B appear at different velocity from most of the region ($+$6.7~km~s$^{-1}$), while IRAS~4C and the surrounding area peak at about 1~km~s$^{-1}$ higher velocity ($+$7.9~km~s$^{-1}$). The observed absorption of the HCN profiles at $+$7.84~km~s$^{-1}$ can be an indicator that IRAS~4C may not be member of IRAS~4 cloud, but probably in foreground. Our analysis is not dependent on the distance though, with the exception of the mass calculation. 
\end{itemize} 

\subsection{Discussion}

Our JCMT data alone have revealed differences between the 3 sources that make us reconsider the true nature of IRAS~4C. The major differences appear in the: a) chemical composition (i.e. the spectrum of IRAS~4C appears less rich than the other two), b) the spatial distribution of some species are more extended towards IRAS~4A and IRAS~4B indicating stronger outflow activity, c) the line profiles suggest weak or non--existent outflow activity in IRAS~4C. 

Moreover, the H$_{2}$CO kinetic temperature analysis reveals that IRAS~4C is characterized by the lowest kinetic temperature ($\sim$12~K), followed by IRAS~4A ($\sim$45~K) and IRAS~4B ($\sim$80~K). HCN/HNC and DCO$^{+}$/HCO$^{+}$ ratios support the temperatures estimated from the H$_{2}$CO analysis. Furthermore, we determine the degree of CO depletion (N(CO)/N(H$_{2}$)) and deuteration (N(DCO$^{+}$)/N(HCO$^{+}$)) for each object. We determine the lowest degree of depletion to be towards IRAS~4B where the deuteration is lower. We report values of 12$\%$ for IRAS~4A source, 3.5$\%$ for IRAS~4B, and an upper limit of 20$\%$ for IRAS~4C. Lastly, we find the [HCN]/[HNC] of $\sim$3.5 for IRAS~4A, $\sim$4.2 for IRAS~4B and $\sim$0.85 for IRAS~4C, being in agreement with the temperatures of the 3 sources.

We clearly observe differences between IRAS~4A, IRAS~4B and IRAS~4C, which may be due to (a) differences in mass and/or luminosity (b) a different orientation 
(edge--on vs face--on) or (c) evolution.

Previous studies report L$_{bol}$ of 4.2~$L_{\odot}$, 1.6~$L_{\odot}$, 0.49~$L_{\odot}$ and M$_{env}$ of 7.75~$M_{\odot}$, 3.66~$M_{\odot}$ and 0.5~$M_{\odot}$ for IRAS~4A, IRAS~4B and IRAS~4C respectively as detemined for a temperature of $\sim$~50~K \citep{Enoch09}. We estimate masses of $\sim$2~$M_{\odot}$ for IRAS~4A, 0.6~$M_{\odot}$ for IRAS~4B and 1.5~$M_{\odot}$ for IRAS~4C, using different temperatures for each source as determined in Sec.~\ref{h2co}. The mass of the envelope shows a linear correlation with the luminosity of CO lines \citep{Irene2013}. The fact that C$^{17}$O emission towards IRAS~4C is stronger than IRAS~4B can be explained by the fact that we find $\sim$3 times higher mass towards IRAS~4C compared to IRAS~4B. The low bolometric luminosity towards IRAS~4C can be the main reason that we observe the weaker lines towards this source. 

\citet{Dunham2008} report L$_{IR}$ $=$ 0.305$\lsun$ for IRAS~4C as integrated for all existing detections between the 1.25--70~$\mu$m that corresponds to an internal luminosity L$_{INT}$ $=$ 0.51$\lsun$ and excludes IRAS~4C from being a VeLLO since its luminosity is $>$ 0.1$\lsun$.

We finally see an emission from a fourth source, that we call IRAS~4--SWC in this work, the nature of which is not clear. IRAS~4--SWC shows line peaks at lower velocity ($\sim5.3$km~s$^{-1}$) and only narrow lines ($\sim$3km~s$^{-1}$) which can be either a projection effect of the outflow, or an indication that IRAS~4--SWC is a separate core. The non-detected emission at mid-- and far-- infrared at this position, while \citet{Winston2010} reports X--ray emission close to it, in combination with the fact that IRAS~4--SWC can be observed mostly in outflow tracers such as CH$_{3}$OH makes it unlike to be a separate core and connects it more with internal shock from the IRAS~4A outflow.

\subsection{Younger age scenario}

We start with investigating the scenario that IRAS~4C is in younger age than the other two objects. 

A lower temperature can be an indication of earlier evolutionary state as soon as the objects of comparison are characterized by the same luminosity. This is not the case for our sources and thus we cannot base our interpretation on the kinetic temperature alone. 
The deuteration ([N$_{2}$D$^{+}$]/[N$_{2}$H$^{+}$]) has been measured from \citet{Crapsi2005a} to be $\sim 16-23$ \% for a prestellar core (e.g. L1544) and $\sim 5-10$ \% for Class 0 objects (e.g. L1521F). Our results suggest that IRAS~4C is not the source with the higher deuteration as expected for younger sources but our big errors towards IRAS~4C (16$\%$) do not allow us to clearly distinguish it from the other two sources. 

The bolometric temperature can be additionally used to probe the various evolutionary stages from prestellar core to protostars \citep[e.g.;][]{Enoch09}. Prestellar cores are characterized by T$_{bol}$ $\sim$ 10--20~K while T$_{bol}$ $<$ 50~K can be used to classify an object as early Class~0. The reported values for IRAS~4A and IRAS~4B are $\sim$30--36~K \citep[e.g.;][]{Kristensen2012,Motte2001} while for IRAS~4C T$_{bol}$ $=$ 26~K \citep{Dunham2008}. \citet{Enoch09} reports a value of $\sim50$K for all 3 objects. The variations between the reported values in different studies make the use of T$_{bol}$ alone not reliable enough to discriminate between the relative states of the sources.

\citet{Gutermuth2008} report IRAS~4C as a deeply embedded source with incomplete IRAC photometry but a relatively bright MIPS 24~$\mu$m. 
IRAS~4A shows compact point--like emission already at 
3.6~$\mu$m and IRAS~4B at 4.5~$\mu$m \citep{Young2015} being associated with shock knots, while IRAS~4C shows a more extended weaker emission at 8~$\mu$m. This information tells us that IRAS~4C possibly contains a compact hydrostatic object that can result in warm dust to emit at 8 and 24~$\mu$m, excluding IRAS~4C from being a pre--stellar core ($>$70~$\mu$m). 

3--D radiation magneto--hydrodynamics simulations that reproduce the stage of a dense core collapse \citep{Machida2008,Tomida2010} show that a slow poorly collimated outflow can exist, without the presence of a protostar, but rather due to the already formed FHSC. These simulations predict a very slow collimated outflow that does not exceed the velocity of 3~km~s$^{-1}$. On the other hand, both theory and observations \citep{Shang07,Arce06}, report highly collimated outflows from YSOs that reach velocities up to a few tens of km~s$^{-1}$. IRAS~4C does not seem to drive a strong outflow, while IRAS~4A and IRAS~4B do show a strong outflow activity (high velocity gas). The orientation plays an important role when it comes to the observed outflow. Line emission from outflow tracers such as CO, is more sensitive to the different orientation of the sources and can affect the strenght of the observed outflow. Thus the observed differences in the outflow activity could be a result of a) orientation and/or b) evolution. 

Although the orientation of the source may influence the interpretation of the outflow emission, it does not affect the estimated values of the degrees of CO depletion and deuteration. C$^{17}$O, DCO$^{+}$ and HCO$^{+}$ are all species found to trace the envelope. Given the fact that the single dish observations cannot resolve a disk structure, we assume that all of the observed emission comes from the bulk of gas tracing the envelope and thus it is independent of the orientation of the sources. The determined values towards IRAS~4C, point towards a colder, less evolved source compared to IRAS~4A and IRAS~4B, either between the prestellar and protostellar phase (FHSC) or a very young Class~0 object at the earliest accretion phase which makes it one of the very few observed cases \citep[e.g.;][]{Andre1999,Belloche06}. 

\subsection{Older age scenario}

We now investigate the scenario that IRAS~4C is in an older age than the other two objects.

\citet{Tobin2015} \& Segura--Cox et al. 2015 (submitted) report strong evidence of an edge on disk--like structure 
around IRAS~4C, although the S/N in that work was not sufficient to test for Keplerian rotation. A disk--structure indicates that IRAS~4C is a more evolved 
object than a FHSC, probably an object in the transition phase between Class~0 and Class~I. An edge on disk--like structure can cause an underestimation of 
the observed T$_{bol}$ by even $\sim$100~K \citep{Launhardt2013}. This would favor the scenario that IRAS~4C is actually a more evolved object with T$_{bol}$$>$70~K, bringing it closer to the late Class~0 to early Class~I stage. This would also be in agreement with the fact that IRAS~4C is the object with the lowest luminosity, which might indicate less strong accreting activity and thus a more evolved object. Another indicator of evolutionary stage is the ratio L$_{smm}$/L$_{bol}$ which is expected to be higher the less evolved the object is, and it is considered to be more accurate than T$_{bol}$. This was found to be 5\% for IRAS~4A, 5.5\% for IRAS~4B and 6.3\% for IRAS~4C \citep{Dunham2008,Sadavoy2014}. \citet{Launhardt2013} report values of 3\%$<$L$_{smm}$/L$_{bol}$$<$7\% for Class~0 and Class~I objects, while \citet{Andre2000} had previously proposed a threshold of $<$0.3\% for Class~I objects. Furthermore they report 10\%$<$L$_{smm}$/L$_{bol}$$<$30\% for starless cores and objects that could possibly host an embedded FHSC. This could limit the classification of IRAS~4C as FHSC, but the reported values for the 3 sources are close enough, making it difficult to distinguish between the Class~0 and Class~I stage. A rough classification would bring IRAS~4C to the Class~0 stage but probably younger than the other two.  

The collimation and bipolarity of an outflow can be used to distinguish Class~0 objects from the more evolved Class~I objects. 
For Class~I objects 
the outflow angle is wider and as we move further to the evolution, the swept--up material is slowing down, since the driving force 
is reducing \citep{Arce06}. Outflows in Class~0 sources have proven to be much stronger than in Class I objects \citep{Bontemps1996}.

Our CARMA observations indicate the presence of a slow outflow ($\sim$2km~s$^{-1}$) which is not characterized 
by a wide angle and it is rather compact. The nebulosity revealed by Spitzer shows a scattered light cone which has an origin at 
the location of the protostar indicating that the most likely direction of the outflow is orthogonal to the plane of the disk. The edge on disk scenario 
can explain the apparent low outflow velocity and thus partially why the line profiles from JCMT are so narrow towards IRAS~4C and deprived of the 
higher velocity activity but it cannot explain that we do not observe spatially an extended outflow. The relatively wide cone--like structure 
to the east of IRAS~4C which is observed in our CO maps from JCMT, could be connected to the outflow activity. There is the possibility that we do not 
observe similar structure to the west of IRAS~4C, due to the strong interaction with the red lobe of IRAS~4A. A more powerful outflow from 4A would sweep the 
red lobe of IRAS~4C away, but this requires a very specific geometry and distance of the sources and the outflows. This is the only scenario that can 
explain the observed outflow by an older age object, but the difference in $V_{\rm LSR}$ argues against it. Also, this explanation cannot be valid if IRAS~4C is a foreground object, 
since there is no way for the outflow lobes to interact.

On the other hand, an outflow parallel to our line of sight could explain a more compact outflow but that would mean that the observed 
velocity is very close to the real velocity and thus very slow, and thus inconsistent with an older age.

\subsection{Conclusions and future direction}

Neither the younger nor the older age scenario seems to be fully consistent with the body of observational data on IRAS 4C. If IRAS~4C is old enough to reveal a disk structure, we cannot fully explain the missing outflow, and the cold nature of the source. If IRAS~4C is so young that cannot drive a powerful outflow, we cannot explain the low L$_{smm}$/L$_{bol}$ and the disk structure that has previously been observed. IRAS~4C appears not to be the strongest candidate among the FHSC candidates. Nevertheless, if all of the candidates located in Perseus \citep[e.g.;][]{Pezzuto2012,Enoch10} will be confirmed, while statistics predict $<$ 0.2, it would be a strong evidence that the lifespan of these objects is $\ge$ 10$^{3}$ years. This can be a result of a lower mass accretion rate ($<$4$\times$10$^{-5}$$M_{\odot}$/yr) or that special conditions apply to Perseus, perhaps a non--stationary accretion. \citet{Tomida2010} predict a lifespan longer than 10$^{4}$ yrs for a protostar of 0.1$M_{\odot}$ adopting an accretion rate of 10$^{-5}$$M_{\odot}$/yr and 
this would improve the situation in Perseus by an order of magnitude.

Applying FHSC and YSO disk models towards IRAS~4C will help us characterizing more accurately this object. In addition, the detailed study of the outflow of the observed FHSCs and VeLLOs may be helpful in understanding the missing link between these early stages of star formation.

\begin{acknowledgements}
The authors thank Mark Krumholz for useful discussions. J.J.T. is supported by grant 639.041.439 from the Netherlands Organisation for Scientific Research (NWO).
\end{acknowledgements}

\bibliographystyle{aa} 
\bibliography{ref}

\begin{appendix}
\section{Line detections towards the 3 sources}

\begin{table*}[ht]
\scriptsize\addtolength{\tabcolsep}{-2.0pt}
\caption{Molecular Line Detections for NGC~1333-IRAS~4A - JCMT data}
\begin{tabular}{c c c c c c c c c c}
\toprule
& & & & \bf Broad & \bf component & & \bf Narrow & \bf component & \\
\hline\hline
Molecule & Transition & Rest Frequency & E$_{up}$ & V$_{lsr}$ & FWHM & T$_{A}^{*}$ & V$_{lsr}$ & FWHM & T$_{A}^{*}$ \\ & & (MHz) & (K) & (km~s$^{-1}$) & (km~s$^{-1}$) & (K) & (km~s$^{-1}$) & (km~s$^{-1}$) & (K) \\
\midrule
HCN & 4-3 & 354505.5 & 42.5 & $5.03\pm0.21$ & $13.7\pm0.5$ & $0.67\pm0.03$ & & & \\
H$^{13}$CN & 4-3 & 345339.8 & 41.4 & $6.4\pm0.7$ & $14\pm2$ & $0.04\pm0.01$ & $7.8\pm0.3$ & $2.8\pm1.0$ & $0.03\pm0.01$ \\
HNC & 4-3 & 362630.3 & 43.5 & $7.66\pm0.05$ & $1.18\pm0.08$ & $1.26\pm0.08$ & & & \\
CO & 3-2 & 345796.0 & 33.2 & $9.2\pm0.8$ & $22.3\pm2$ & $2.6\pm0.2$ & $6.7\pm0.1$ & $1.5\pm0.2$ & $6.1\pm0.7$ \\
$^{13}$CO & 3-2 & 330588.0 & 31.7 & $7.3\pm0.9$ & $10\pm3$ & $0.4\pm0.1$& $7.63\pm0.05$& $2.1\pm0.2$ & $3.3\pm0.2$\\
C$^{17}$O & 3-2 & 337061.1 & 32.4 & $7.65\pm0.02$ & $1.52\pm0.06$ & $0.70\pm0.02$ & & & \\
HCO$^{+}$ & 4-3 & 356734.1 & 42.8 & $8.87\pm0.55$ & $10.2\pm1.5$ & $0.71\pm0.09$ & $7.41\pm0.04$ & $1.21\pm0.08$ & $3.43\pm0.19$ \\
H$^{13}$CO$^{+}$ & 4-3 & 346998.3 & 41.6 & $7.64\pm0.02$ & $1.24\pm0.04$ & $0.87\pm0.02$ & & & \\
H$_{2}$$^{13}$CO & 5(1,5)-4(1,4) & 343325.7 & 61.3 & $6.9\pm0.4$ & $3\pm1$ & $0.06\pm0.2$ & & & \\
H$_{2}$$^{13}$CO & 5(2,3)-4(2,2) & 356176.2 & 98.5 & $8.0\pm0.2$ & $2.6\pm0.7$ & $0.06\pm0.01$ & & & \\
CS & 7-6 & 342882.8 & 65.8 & $6.53\pm0.09$ & $11.77\pm0.25$ & $0.69\pm0.01$ & $7.53\pm0.02$ & $1.26\pm0.12$ & $0.87\pm0.05$ \\
C$^{34}$S & 7-6 & 337396.7 & 50.2 & $7.1\pm0.2$ & $3.2\pm0.5$ & $0.16\pm0.02$ & & & \\
H$_{2}$CS & 10(1,10)-9(1,9) & 338083.2 & 102.4 & $7.54\pm0.52$ & $7.0\pm1.4$ & $0.06\pm0.01$ & & & \\
H$_{2}$CS & 10(2,9)-9(2,8) & 343322.1 & 143.3 & $4.3\pm0.7$ & $6.0\pm1.7$ & $0.04\pm0.01$ & & & \\
H$_{2}$CS & 10(1,9)-9(1,8) & 348534.4 & 105.2 & $8.0\pm0.6$ & $5.1\pm1.5$ & $0.06\pm0.01$ & & & \\
OCS & 28-27 & 340449.2 & 237.0 & $7.6\pm0.2$ & $3.8\pm0.6$ & $0.054\pm0.007$ & & & \\
CH$_{3}$OH & 11(1,10)-11(0,11) & 331502.3 & 169.0 & $4.6\pm0.6$ & $10.0\pm1.4$ & $0.06\pm0.01$ & $7.07\pm0.47$ & $2.29\pm1.34$ & $0.02\pm0.01$ \\
CH$_{3}$OH & 7(1,7)-6(1,6) & 335582.0 & 79.0 & $5.1\pm1.4$ & $9.2\pm3.0$ & $0.04\pm0.04$ & $7.8\pm1.5$ & $2.7\pm5.2$ & $0.04\pm0.04$ \\
CH$_{3}$OH & 7(0,7)-6(0,6) & 338124.5 & 78.1 & $5.5\pm0.4$ & $10.2\pm0.9$ & $0.12\pm0.01$ & $7.8\pm0.2$ & $1.0\pm0.5$ & $0.10\pm0.03$ \\
CH$_{3}$OH & 7(-1,7)-6(-1,6) & 338344.6 & 70.6 & $5.3\pm0.1$ & $10.2\pm0.2$ & $0.27\pm0.08$ & $7.61\pm0.05$ & $2.16\pm0.14$ & $0.25\pm0.01$ \\
CH$_{3}$OH & 7(-1,7)-6(-1,6)++ & 338408.7 & 65.0 & $5.4\pm0.2$& $10.3\pm0.4$ & $0.34\pm0.02$ & $7.48\pm0.08$ & $2.04\pm0.22$ & $0.29\pm0.02$ \\
CH$_{3}$OH & 7(2,6)-6(2,5)-- & 338512.9 & 102.7 & $7.2\pm0.7$ & $7.2\pm1.8$ & $0.04\pm0.01$ & & & \\
CH$_{3}$OH & 7(3,4)- 6(3,3) & 338583.2 & 112.7 & $7.6\pm0.2$ & $2.6\pm0.6$ & $0.05\pm0.01$ & & & \\
CH$_{3}$OH & 7(3,5)-6(3,4)++ & 338540.8 & 114.8 & $5.9\pm0.2$ & $4.5\pm0.5$ & $0.07\pm0.01$ & & & \\
CH$_{3}$OH & 7(1,6)-6(1,5) & 338615.0 & 86.0 & $7.8\pm0.3$ & $10.2\pm0.9$ & $0.08\pm0.01$ & $7.7\pm0.1$ & $1.7\pm0.4$ & $0.05\pm0.01$ \\
CH$_{3}$OH & 7(2,5)-6(2,4) & 338721.7 & 87.3 & $6.1\pm0.2$ & $8.5\pm0.5$ & $0.14\pm0.01$ & $7.9\pm0.2$ & $2.0\pm0.6$ & $0.06\pm0.01$ \\
CH$_{3}$OH & 7(-2,5)-6(-2,4) & 338722.9 & 90.9 & $7.1\pm0.2$ & $8.5\pm0.5$ & $0.14\pm0.01$ & $8.9\pm0.2$ & $2.1\pm0.6$ & $0.06\pm0.01$ \\
CH$_{3}$OH & 7(1,6)-6(1,5)-- & 341415.6 & 80.1 & $5.2\pm0.3$ & $8.2\pm0.6$ & $0.11\pm0.01$ & $7.2\pm0.3$ & $2.9\pm1.0$ & $0.05\pm0.02$ \\
CH$_{3}$OH & 13(1,12)-13(0,13)-+ & 342729.8 & 227.5 & $7.0\pm1.0$ & $8.5\pm3.1$ & $0.03\pm0.01$ & & & \\
CH$_{3}$OH & 4(0,4)-3(-1,3) & 350687.7 & 36.3 & $4.8\pm0.7$ & $9.0\pm1.4$ & $0.09\pm0.02$ & $6.9\pm0.2$ & $2.7\pm0.6$ & $0.14\pm0.02$ \\
CH$_{3}$OH & 1(1,1)-0(0,0)++ & 350905.1 & 16.8 & $4.4\pm1.2$ & $7.3\pm2.5$ & $0.11\pm0.03$ & $6.9\pm0.4$ & $2.2\pm1.2$ & $0.13\pm0.05$ \\
CH$_{3}$OH & 4(1,3)-3(0,3) & 358605.8 & 44.3 & $5.2\pm0.5$ & $8.7\pm0.9$ & $0.10\pm0.01$ & $7.4\pm0.1$ & $2.2\pm0.4$ & $0.12\pm0.02$ \\
CH$_{3}$OH & 8(1,7)-7(2,5) & 361852.2 & 104.6 & $6.8\pm0.4$ & $4.7\pm1.0$ & $0.03\pm0.001$ & & & \\
CH$_{3}$OH & 7(2,5)-6(1,5) & 363739.8 & 87.2 & $7.5\pm0.9$ & $11.4\pm4.4$ & $0.05\pm0.02$ & $7.3\pm0.3$ & $2.9\pm1.3$ & $0.06\pm0.02$ \\
H$_{2}$CO & 5(1,5)-4(1,4) & 351768.6 & 62.4 & $6.0\pm0.1$ & $9.0\pm0.4$ & $0.71\pm0.03$ & $7.43\pm0.03$ & $1.41\pm0.08$ & $1.16\pm0.05$ \\
H$_{2}$CO & 5(0,5)-4(0,4) & 362736.0 & 52.3 & $7.16\pm0.09$ & $9.1\pm0.2$ & $0.28\pm0.01$ & $7.81\pm0.02$ & $1.40\pm0.07$ & $0.53\pm0.04$ \\
H$_{2}$CO & 5(2,4)-4(2,3) & 363945.9 & 99.5 & $6.49\pm0.05$ & $11.1\pm0.1$ & $0.08\pm0.01$ & $8.01\pm0.01$ & $1.94\pm0.03$ & $0.12\pm0.01$ \\
H$_{2}$CO & 5(2,3)-4(2,2) & 365363.4 & 99.7 & $7.34\pm0.08$ & $8.5\pm0.2$ & $0.12\pm0.02$ & $8.25\pm0.02$ & $0.91\pm0.05$ & $0.15\pm0.03$ \\
H$_{2}$CO & 5(3,3)-4(3,2) & 364275.1 & 158.4 & $7.0\pm3.1$ & $11.0\pm4.8$ & $0.09\pm0.03$ & $8.3\pm0.2$ & $1.8\pm0.7$ & $0.08\pm0.02$ \\
H$_{2}$CO & 5(3,2)-4(3,1) & 364288.9 & 158.4 & $5.8\pm1.9$ & $12.4\pm3.4$ & $0.09\pm0.01$ & $7.6\pm0.2$ & $1.4\pm0.5$ & $0.08\pm0.02$ \\
H$_{2}$CO & 5(4,2)-4(4,1) & 364103.2 & 240.7 & $9.0\pm0.7$ & $5.6\pm1.9$ & $0.04\pm0.01$ & & & \\
DCO$^{+}$ & 5-4 & 360169.8 & 51.9 & $7.56\pm0.02$ & $1.11\pm0.07$ & $0.45\pm0.02$ & & & \\ 
SO $^{3}\Sigma$ & 7(8)-6(7) & 340714.2 & 81.2 & $5.2\pm0.6$ & $14.3\pm1.5$ & 0.10$\pm$0.01 & & & \\
SO $^{3}\Sigma$ & 8(8)-7(7) & 344310.6 & 87.5 & $5.2\pm0.3$ & $12.5\pm0.8$ & $0.118\pm0.006$ & & & \\
SO $^{3}\Sigma$ & 8(9)-7(8) & 346528.5 & 78.8 & $5.4\pm0.3$ & $14.5\pm0.9$ & $0.27\pm0.01$ & & & \\
$^{33}$SO & 10(11)-10(10) & 332014.3 & 141.5 & $5.5\pm0.1$ & $4.0\pm0.2$ & $0.187\pm0.008$ & & & \\
SO$_{2}$ & 4(3,1)-3(2,2) & 332505.2 & 31.3 & $7.7\pm0.7$ & $5\pm1$ & $0.04\pm0.01$ & & & \\
C$_{2}$H & N=4-3, J=9/2-7/2, F= 4-3 & 349337.5 & 41.9 & $7.8\pm0.2$ & $4.9\pm0.5$ & $0.09\pm0.01$ & & & \\
C$_{2}$H & N=4-3, J=7/2-5/2, F= 4-3 & 349399.3 & 41.9 & $8.0\pm0.7$ & $5.4\pm1.9$ & $0.06\pm0.01$ & & & \\
CN & N=3-2, J=5/2-3/2, F=7/2-5/2 & 340031.5 & 32.6 & $6.9\pm1$ & $5\pm3$ & $0.09\pm0.03$ & & & \\
CN & N=3-2, J=7/2-5/2, F=9/2-7/2 & 340247.8 & 32.7 & $7.87\pm0.05$ & $1.5\pm0.1$ & $0.38\pm0.02$ & & & \\
HDCO & 5(1,4)-4(1,3) & 335096.9 & 56.2 & $7.5\pm0.2$ & $1.1\pm0.3$ & $0.09\pm0.02$ & & & \\
HDCO & 6(1,6)-5(1,5) & 369763.5 & 70.1 & $7.6\pm0.2$ & $1.0\pm0.4$ & $0.10\pm0.03$ & & & \\
D$_{2}$CO & 6(0,6)-5(0,5) & 342522.1 & 58.1 & $7.7\pm0.2$ & $2.7\pm0.8$ & $0.08\pm0.01$ & & & \\
SiO & 8-7 & 347330.6 & 75.0 & $-$0.5$\pm$1.7 & 24$\pm$6 & 0.04$\pm$0.01 & 4.4$\pm$0.5 & 4.7$\pm$1.5 & 0.05$\pm0.01$ \\
N$_{2}$H$^{+}$ & 4-3 & 372672.5 & 44.7 & $7.52\pm0.08$ & $3.47\pm0.34$ & $0.41\pm0.78$ & $7.39\pm0.05$ & $1.0\pm0.3$ & $1.8\pm0.7$ \\
\bottomrule
\label{JCMT_lines_1}
\end{tabular}
\label{JCMT_lines_1}
\end{table*}

\begin{table*}[htbp]
\scriptsize\addtolength{\tabcolsep}{-2.0pt}
\caption{Molecular Line Detections for NGC~1333-IRAS~4B - JCMT data}
\begin{tabular}{c c c c c c c c c c}
\toprule
& & & & \bf Broad & \bf component & & \bf Narrow & \bf component & \\
\hline\hline
Molecule & Transition & Rest Frequency & E$_{up}$ & V$_{lsr}$ & FWHM & T$_{A}^{*}$ & V$_{lsr}$ & FWHM & T$_{A}^{*}$ \\ & & (MHz) & (K) & (km~s$^{-1}$) & (km~s$^{-1}$) & (K) & (km~s$^{-1}$) & (km~s$^{-1}$) & (K) \\
\hline\hline
HCN & 4-3 & 354505.5 & 42.5 & $8.0\pm0.2$ & $11.5\pm0.5$ & $1.04\pm0.05$ & $7.11\pm0.04$ & $0.9\pm0.8$ & $1.2\pm1$ \\
H$^{13}$CN & 4-3 & 345339.8 & 41.4 & $8.4\pm0.3$ & $7.1\pm0.8$ & $0.08\pm0.01$ & & & \\
HNC & 4-3 & 362630.3 & 43.5 & $7.8\pm0.2$ & $4.5\pm0.6$ & $0.12\pm0.02$ & $7.67\pm0.02$ & $1.09\pm0.04$ & $0.89\pm0.03$ \\
CO & 3-2 & 345796.0 & 33.2 & $9.3\pm0.4$ & $14.6\pm1.0$ & $1.2\pm0.1$ & $6.78\pm0.04$ & $1.8\pm0.1$ & $4.2\pm0.2$ \\
$^{13}$CO & 3-2 & 330587.9 & 31.7 & $7.90\pm0.03$ & $2.56\pm0.08$ & $2.8\pm0.1$ & & & \\
C$^{17}$O & 3-2 & 337061.1 & 32.4 & $7.4\pm0.1$ & $1.8\pm0.1$ & $0.37\pm0.03$ & & & \\
HCO$^{+}$ & 4-3 & 356734.1 & 42.8 & $9.5\pm0.4$ & $9.9\pm0.9$ & $0.45\pm0.04$ & $7.47\pm0.03$ & $1.36\pm0.07$ & $2.1\pm0.1$ \\
H$^{13}$CO$^{+}$ & 4-3 & 346998.3 & 41.6 & $7.70\pm0.05$ & $1.26\pm0.09$ & $0.35\pm0.02$ & & & \\
H$_{2}$$^{13}$CO & 5(1,5)-4(1,4) & 343325.7 & 61.3 & $8.3\pm0.2$ & $1.6\pm0.3$ & $0.11\pm0.02$ & & & \\
CS & 7-6 & 342882.8 & 65.8 & $8.0\pm0.1$ & $10.5\pm0.2$ & $0.63\pm0.02$ & $7.86\pm0.03$ & $2.6\pm0.1$ & $0.77\pm0.03$ \\
CH$_{3}$OH & 11(1,10)-11(0,11) & 331502.4 & 169.0 & $8.5\pm0.5$ & $9.6\pm2.2$ & $0.08\pm0.03$ & $8.0\pm0.1$ & $3.5\pm0.6$ & $0.15\pm0.03$ \\
CH$_{3}$OH & 7(1,7)-6(1,6) & 335582.0 & 79.0 & $8.3\pm0.2$ & $5.3\pm0.4$ & $0.18\pm0.02$ & $7.4\pm0.1$ & $1.3\pm0.3$ & $0.14\pm0.03$ \\
CH$_{3}$OH & 12(1,11)-12(0,12) & 336865.1 & 197.1 & $8.5\pm0.1$ & $4.8\pm0.4$ & $0.14\pm0.01$ & & & \\
CH$_{3}$OH & 7(0,7)-6(0,6) & 338124.5 & 78.1 & $9.3\pm0.4$ & $7.2\pm0.7$ & $0.15\pm0.02$ & $7.78\pm0.09$ & $2.2\pm0.3$ & $0.27\pm0.03$ \\
CH$_{3}$OH & 7(-1,6)-6(-1,6) & 338344.6 & 70.5 & $8.8\pm0.1$ & $7.0\pm0.4$ & $0.34\pm0.03$ & $7.81\pm0.04$ & $2.2\pm0.1$ & $0.58\pm0.03$ \\
CH$_{3}$OH & 7(6,2)-6(6,1) - & 338404.6 & 243.8 & $5.2\pm0.2$ & $7.3\pm0.4$ & $0.39\pm0.04$ & $4.16\pm0.05$ & $2.1\pm0.2$ & $0.67\pm0.04$ \\
CH$_{3}$OH & 7(-1,7)-6(-1,6)++ & 338408.7 & 65.0 & $8.8\pm0.2$ & $7.3\pm0.4$ & $0.38\pm0.04$ & $7.79\pm0.05$ & $2.2\pm0.1$ & $0.67\pm0.04$ \\
CH$_{3}$OH & 7(2,6)-6(2,5)-- & 338512.9 & 102.7 & $8.5\pm0.1$ & $3.6\pm0.2$ & $0.13\pm0.01$ & & & \\
CH$_{3}$OH & 7(3,5)-6(3,4)++ & 338540.8 & 114.8 & $7.5\pm0.1$ & $5.0\pm0.3$ & $0.15\pm0.01$ & & & \\
CH$_{3}$OH & 7(3,4)- 6(3,3) & 338583.2 & 112.7 & $8.6\pm0.4$ & $2.9\pm0.8$ & $0.06\pm0.01$ & & & \\
CH$_{3}$OH & 7(1,6)-6(1,5) & 338615.0 & 86.0 & $9.2\pm0.4$ & $6.8\pm0.8$ & $0.12\pm0.02$ & $7.9\pm0.1$ & $1.9\pm0.4$ & $0.17\pm0.03$ \\
CH$_{3}$OH & 7(2,5)-6(2,4)++ & 338639.9 & 102.7 & $8.4\pm0.2$ & $4.8\pm0.6$ & $0.08\pm0.01$ & & & \\
CH$_{3}$OH & 7(2,5)-6(2,4) & 338721.6 & 87.3 & $11.8\pm0.8$ & $4.1\pm1.4$ & $0.09\pm0.02$ & $7.7\pm0.16$ & $3.8\pm0.2$ & $0.45\pm0.02$ \\
CH$_{3}$OH & 7(1,6)-6(1,5)-- & 341415.6 & 80.1 & $8.9\pm0.2$ & $6.3\pm0.4$ & $0.18\pm0.02$ & $7.78\pm0.05$ & $2.1\pm0.2$ & $0.31\pm0.02$ \\
CH$_{3}$OH & 13(1,12)-13(0,13)-+ & 342729.8 & 227.5 & $8.6\pm0.3$ & $6.3\pm0.6$ & $0.11\pm0.02$ & $7.3\pm0.3$ & $1.9\pm0.8$ & $0.06\pm0.02$ \\
CH$_{3}$OH & 4(0,4)-3(-1,3) & 350687.7 & 36.3 & $8.3\pm0.2$ & $5.7\pm0.6$ & $0.15\pm0.02$ & $7.49\pm0.07$ & $1.7\pm0.2$ & $0.23\pm0.03$ \\
CH$_{3}$OH & 1(1,1)-0(0,0)++ & 350905.1 & 16.8 & $8.8\pm0.3$ & $6.0\pm0.5$ & $0.16\pm0.02$ & $7.48\pm0.05$ & $1.3\pm0.3$ & $0.28\pm0.03$ \\
CH$_{3}$OH & 13(0,13)-12(1,12)++ & 355603.1 & 211.0 & $8.6\pm0.3$ & $5.7\pm0.8$ & $0.14\pm0.02$ & & & \\
CH$_{3}$OH & 4(1,3)-3(0,3) & 358605.8 & 44.3 & $8.0\pm0.2$ & $4.7\pm0.7$ & $0.18\pm0.06$ & $7.8\pm0.1$ & $1.7\pm0.4$ & $0.20\pm0.06$ \\
CH$_{3}$OH & 11(0,11)-10(1,9) & 360848.9 & 166.0 & $8.3\pm0.2$ & $6.6\pm0.7$ & $0.05\pm0.01$ & $8.0\pm0.2$ & $1.2\pm1.0$ & $0.02\pm0.01$ \\
CH$_{3}$OH & 8(1,7)-7(2,5) & 361852.2 & 104.6 & $7.9\pm0.6$ & $7.2\pm1.9$ & $0.05\pm0.01$ & $7.5\pm0.1$ & $0.9\pm3.2$ & $0.08\pm0.30$ \\
CH$_{3}$OH & 7(2,5)-6(1,5) & 363739.8 & 87.2 & $8.7\pm0.2$ & $6.8\pm0.6$ & $0.16\pm0.02$ & $8.06\pm0.06$ & $1.7\pm0.2$ & $0.25\pm0.02$ \\
H$_{2}$CO & 5(1,5)-4(1,4) & 351768.6 & 62.4 & $8.7\pm0.2$ & $8.2\pm0.5$ & $0.66\pm0.07$ & $7.60\pm0.03$ & $2.6\pm0.1$ & $1.81\pm0.07$ \\
H$_{2}$CO & 5(0,5)-4(0,4) & 362736.0 & 52.3 & $7.9\pm0.03$ & $2.2\pm0.1$ & $1.17\pm0.04$ & $7.86\pm0.03$ & $2.25\pm0.09$ & $1.17\pm0.04$ \\
H$_{2}$CO & 5(2,4)-4(2,3) & 363945.8 & 99.5 & $8.4\pm0.1$ & $5.1\pm0.4$ & $0.27\pm0.04$ & $7.94\pm0.06$ & $1.9\pm0.2$ & $0.33\pm0.04$ \\
H$_{2}$CO & 5(2,3)-4(2,2) & 365363.4 & 99.7 & $9.1\pm0.3$ & $9.0\pm0.9$ & $0.13\pm0.02$ & $8.33\pm0.04$ & $2.6\pm0.2$ & $0.48\pm0.02$ \\
H$_{2}$CO & 5(3,3)-4(3,2) & 364275.1 & 158.4 & $8.7\pm0.5$ & $7.5\pm1.1$ & $0.18\pm0.04$ & $8.16\pm0.08$ & $2.5\pm0.3$ & $0.36\pm0.04$ \\
H$_{2}$CO & 5(3,2)-4(3,1) & 364288.9 & 158.4 & $9.4\pm0.9$ & $7.9\pm2.0$ & $0.16\pm0.04$ & $8.16\pm0.07$ & $2.7\pm0.3$ & $0.40\pm0.05$ \\
H$_{2}$CO & 5(4,2)-4(4,1) & 364103.25 & 240.7 & $8.5\pm0.2$ & $3.9\pm0.4$ & $0.11\pm0.01$ & & & \\
DCO$^{+}$ & 5-4 & 360169.8 & 51.9 & $7.65\pm0.09$ & $0.8\pm0.4$ & $0.6\pm0.3$ & & & \\ 
SO $^{3}\Sigma$ & 3(3)-3(2) & 339341.5 & 25.5 & $8.5\pm0.1$ & $1.3\pm0.2$ & $0.16\pm0.02$ & & & \\ 
SO $^{3}\Sigma$ & 7(8)-6(7) & 340714.2 & 81.2 & $8.2\pm0.4$ & $7.8\pm1.1$ & $0.07\pm0.01$ & $7.1\pm0.2$ & $1.3\pm0.6$ & $0.06\pm0.02$\\
SO $^{3}\Sigma$ & 8(8)-7(7) & 344310.61 & 87.5 & $9.4\pm0.5$ & $10.1\pm1.1$ & $0.06\pm0.01$ & $7.6\pm0.1$ & $1.9\pm0.6$ & $0.06\pm0.02$ \\
SO $^{3}\Sigma$ & 9(8)-8(7) & 346528.5 & 78.8 & $9.5\pm0.4$ & $12.7\pm0.9$ & $0.10\pm0.01$ & $7.8\pm0.1$ &$2.4\pm0.3$ & $0.13\pm0.01$\\
SO$_{2}$ & 4(3,1)-3(2,2) & 332505.2 & 31.3 & $7.9\pm0.2$ & $1.7\pm0.5$ & $0.08\pm0.01$ & & & \\
C$_{2}$H & N=4-3, J=9/2-7/2, F= 4-3 & 349337.5 & 41.9 & $7.2\pm0.2$ & $2.6\pm0.5$ & $0.10\pm0.02$ & & & \\
C$_{2}$H & N=4-3, J=7/2-5/2, F= 4-3 & 349399.3 & 41.9 & $6.8\pm0.4$ & $3.7\pm1.0$ & $0.06\pm0.01$ & & & \\
CN & N=3-2, J=5/2-3/2, F=7/2-5/2 & 340031.5 & 32.6 & $9.7\pm1$ & $7.6\pm3$ & $0.06\pm0.02$ & & & \\
CN & N=3-2, J=7/2-5/2, F=9/2-7/2 & 340247.8 & 32.7 & $7.7\pm0.2$ & $1.9\pm0.4$ & $0.23\pm0.03$ & & & \\
HDCO & 5(1,4)-4(1,3) & 335096.8 & 56.2 & $7.6\pm0.1$ & $1.8\pm0.3$ & $0.17\pm0.02$ & & & \\
HDCO & 6(1,6)-5(1,5) & 369763.5 & 70.1 & $7.1\pm0.2$ & $1.7\pm0.6$ & $0.13\pm0.03$ & & & \\
D$_{2}$CO & 6(0,6)-5(0,5) & 342522.1 & 58.1 & $7.8\pm0.2$ & $1.6\pm0.4$ & $0.10\pm0.03$ & & & \\
SiO & 8-7 & 347330.6 & 75.0 & $8.6\pm1.0$ & $30.0\pm3.0$ & $0.08\pm0.01$ & & & \\
N$_{2}$H$^{+}$ & 4-3 & 372672.5 & 44.7 & $7.3\pm0.1$ & $3.5\pm0.4$ & $0.4\pm0.1$ & $7.3\pm0.02$ & $1.0\pm0.1$ & $1.8\pm0.1$ \\

\bottomrule
\end{tabular}
\label{JCMT_lines_2}
\end{table*}

\begin{table*}[tp]
\scriptsize\addtolength{\tabcolsep}{-2.0pt}
\caption{Molecular Line Detections for NGC~1333-IRAS~4C - JCMT data}
\begin{tabular}{c c c c c c c c c c}
\toprule
& & & & \bf Broad & \bf component & & \bf Narrow & \bf component & \\
\hline\hline
Molecule & Transition & Rest Frequency & E$_{up}$ & V$_{lsr}$ & FWHM & T$_{A}^{*}$ & V$_{lsr}$ & FWHM & T$_{A}^{*}$ \\ & & (MHz) & (K) & (km~s$^{-1}$) & (km~s$^{-1}$) & (K) & (km~s$^{-1}$) & (km~s$^{-1}$) & (K) \\
\hline\hline
HCN & 4-3 & 354505.5 & 42.5 & $8.35\pm0.06$ & $1.2\pm0.1$ & $0.21\pm0.02$ & & & \\
HNC & 4-3 & 362630.3 & 43.5 & $8.58\pm0.05$ & $1.15\pm0.09$ & $0.42\pm0.03$ & & & \\
CO & 3-2 & 345796.0 & 33.2 & $8.7\pm0.5$ & $5.1\pm0.9$ & $2.6\pm0.4$ & $6.8\pm0.1$ & $1.3\pm0.3$ & $4.1\pm0.9$ \\
$^{13}$CO & 3-2 & 330588.0 & 31.7 & $8.2\pm1.2$ & $1.9\pm0.1$ & $2.9\pm0.1$ & & & \\
C$^{17}$O & 3-2 & 337061.1 & 32.4 & $8.27\pm0.02$ & $0.96\pm0.05$ & $0.69\pm0.01$ & & & \\
HCO$^{+}$ & 4-3 & 356734.1 & 42.8 & $8.5\pm0.03$ & $1.52\pm0.07$ & $0.99\pm0.04$ & & & \\
H$^{13}$CO$^{+}$ & 4-3 & 346998.3 & 41.6 & $8.44\pm0.06$ & $1.2\pm0.1$ & $0.20\pm0.01$ & & & \\
CS & 7-6 & 342882.8 & 65.8 & $8.5\pm0.1$ & $1.1\pm0.3$ & $0.35\pm0.08$ & & & \\
CH$_{3}$OH & 4(0,4)-3(-1,3) & 350687.7 & 36.3 & $7.6\pm0.3$ & $3.4\pm0.6$ & $0.09\pm0.01$ & & & \\
H$_{2}$CO & 5(1,5)-4(1,4) & 351768.6 & 62.4 & $8.31\pm0.04$ & $1.3\pm0.1$ & $0.40\pm0.03$ & & & \\
H$_{2}$CO & 5(0,5)-4(0,4) & 362736.0 & 52.3 & $8.72\pm0.08$ & $1.4\pm0.2$ & $0.20\pm0.02$ & & & \\
DCO$^{+}$ & 5-4 & 360169.8 & 51.9 & $8.4\pm0.5$ & $0.9\pm0.5$ & $0.2\pm0.1$ & & & \\ 
SO $^{3}\Sigma$ & 3(3)-3(2) & 339341.5 & 25.5 & $8.2\pm0.3$ & $1.8\pm0.8$ & $0.09\pm0.03$ & & & \\ 
SO $^{3}\Sigma$ & 7(8)-6(7) & 340714.2 & 81.2 & $8.7\pm0.2$ & $2.0\pm0.4$ &$0.06\pm0.01$ & & & \\
SO $^{3}\Sigma$ & 8(8)-7(7) & 344310.6 & 87.5 & $8.8\pm0.8$ & $0.8\pm1.8$ & $0.1\pm0.2$ & & & \\
SO $^{3}\Sigma$ & 9(8)-8(7) & 346528.5 & 78.8 & $8.6\pm0.1$ & $1.9\pm0.1$ & $0.11\pm0.01$ & & & \\
C$_{2}$H & N=4-3, J=9/2-7/2, F= 4-3 & 349337.5 & 41.9 & $7.9\pm0.2$ & $1.9\pm0.4$ & $0.21\pm0.03$ & & & \\
C$_{2}$H & N=4-3, J=7/2-5/2, F= 4-3 & 349399.3 & 41.9 & $7.7\pm0.1$ & $1.9\pm0.3$ & $0.17\pm0.02$ & & & \\
CN & N=3-2, J=5/2-3/2, F=7/2-5/2 & 340031.5 & 32.6 & $9.2\pm0.2$ & $2.4\pm0.5$ & $0.09\pm0.01$ & & & \\
CN & N=3-2, J=7/2-5/2, F=9/2-7/2 & 340247.8 & 32.7 & $8.7\pm0.1$ & $1.4\pm0.3$ & $0.22\pm0.04$ & & & \\
N$_{2}$H$^{+}$ & 4-3 & 372672.5 & 44.7 & $8.1\pm0.1$ & $0.9\pm0.2$ & $0.55\pm0.05$ & & & \\
H$_{2}$D$^{+}$ & 1(1,0)-1(1,1) & 372421.4 & 104.2 & $8.4\pm1.1$ & $1.7\pm1.2$ & $0.08\pm0.03$ & & & \\
\bottomrule
\end{tabular}
\label{JCMT_lines_3}
\end{table*}

\end{appendix}
\end{document}